\documentclass[iop]{emulateapj}
\usepackage{graphicx,xcolor,verbatim}
\usepackage{appendix}
\usepackage{lineno}
\pdfoutput=1

\slugcomment{}

\shorttitle{Star formation efficiency in a turbulent galaxy}

\definecolor{burgundy}{rgb}{0.5, 0.0, 0.13}

\begin{document}

\title{Extreme Variation in Star Formation Efficiency Across a Compact, Starbursting Disk Galaxy}


\author{Fisher, D.B.\altaffilmark{1,2}, Bolatto, A.D.\altaffilmark{3}, Glazebrook, K\altaffilmark{1,2}, Obreschkow, D\altaffilmark{4}, Abraham, R.G.\altaffilmark{5}, Kacprzak, G.G.\altaffilmark{1,2},   \& Nielsen, N.M.\altaffilmark{1,2}}


\altaffiltext{1}{Centre for Astrophysics and Supercomputing, Swinburne University of Technology, P.O. Box 218, Hawthorn, VIC 3122, Australia}
\altaffiltext{2}{ARC Centre of Excellence for All Sky Astrophysics in 3 Dimensions (ASTRO 3D)}
\altaffiltext{3}{Laboratory of Millimeter Astronomy, University of Maryland, College Park, MD 29742}
\altaffiltext{4}{International Centre for Radio Astronomy Research (ICRAR), M468, University of Western Australia, 35 Stirling Hwy, Crawley, WA 6009, Australia}
\altaffiltext{5}{Department of Astronomy \& Astrophysics, University of Toronto, 50 St. George St., Toronto, ON M5S 3H8, Canada}


\begin{abstract}
We report on the internal distribution of star formation efficiency in IRAS~08339+6517 (hereafter IRAS08), using $\sim$200~pc resolution CO(2-1) observations from NOEMA. The molecular gas depletion time changes by 2 orders-of-magnitude from disk-like values in the outer parts to less than 10$^8$~yr inside the half-light radius. This translates to a star formation efficiency per free-fall time that also changes by 2 orders-of-magnitude, reaching 50-100\%, different than local spiral galaxies and typical assumption of constant, low star formation efficiencies. Our target is a compact, massive disk galaxy that has SFR 10$\times$ above the $z=0$ main-sequence; Toomre $Q\approx0.5-0.7$ and high gas velocity dispersion  ($\sigma_{mol}\approx 25$~km~s$^{-1}$). We find that IRAS08 is similar to other rotating, starburst galaxies from the literature in the resolved $\Sigma_{SFR}\propto\Sigma_{mol}^N$ relation. By combining resolved literature studies we find that distance from the main-sequence is a strong indicator of the Kennicutt-Schmidt powerlaw slope, with slopes of $N\approx1.6$ for starbursts from 100-10$^4$~M$_{\odot}$~pc$^{-2}$. Our target is consistent with a scenario in which violent disk instabilities drive rapid inflows of gas. It has low values of Toomre-$Q$, and also at all radii the inflow timescale of the gas is less than the depletion time, which is consistent with the flat metallicity gradients in IRAS08.  We consider these results in light of popular star formation theories, in general observations of IRAS08 find the most tension with theories in which star formation efficiency is a constant. Our results argue for the need of high spatial resolution CO observations are a larger number of similar targets.    
\end{abstract}

\keywords{galaxies: evolution --- galaxies: star formation --- galaxies: starburst ---galaxies: individual(\objectname{IRAS08339+6517})
}


\section{Introduction}
The connection between gas and star formation rate in galaxies, either measured as the depletion time or the star formation efficiency, provides direct test to star formation theories and is a direct input to models of galaxy evolution \citep[for review][]{kennicuttevans2012arxiv,Tacconi2020arxiv,hodge2020arxiv}. The last decade has had a wealth of such studies in large disk galaxies of the local Universe \citep[e.g.][]{bigiel2008,leroy2008,rahman2012,leroy2013,fisher2013,Utomo2017,Leroy2017}. Local Universe studies find that in the main bodies of disks (R$_{gal}>0.1R_{25}$) the ratio of molecular gas to star formation rate (SFR) surface density, the so-called depletion time, is consistently found to be $t_{dep}\sim$1-2~Gyr with statistically significant scatter at the 0.3~dex level. This behavior extends into atomic gas dominated regions of galaxies \citep{Schruba2011}. \cite{Utomo2017} reports a trend toward lower $t_{dep}$ in the central 10\% of the galaxy disk in 14 of 54 galaxies from the CARMA-EDGE survey, yet this variation is rarely larger than a factor of $\sim2-3\times$.  
In general the picture of star formation in the disks of $z=0$ is for the most part a regular process, with variation in molecular gas depletion time typically no greater than the 0.3~dex level. 


%

We know much less about the resolved relationship between gas and SFR surface density in starbursting systems, which are typically found in either advanced stage mergers or turbulent disks of the $z>1$ Universe. Pioneering observations have been made of gas mostly in the brightest star-bursting systems at $z>1$ \citep[e.g.][]{genzel2013,hodge2015,swinbank2011,chen2017aless,tadaki2018,Sharon2019}. Unlike like with local spirals, the combined data set of these individual target studies does not show a simple single power-law in the $\sim$1~kpc resolved relationship between $\Sigma_{SFR}$ and $\Sigma_{mol}$, nor is $t_{dep}$ always found to be constant inside $z>1$ disks \citep[e.g.][]{hodge2015,tadaki2018}. The few observations we have imply a far more complex picture at the peak of cosmic star formation. Recently, kiloparsec-scale resolved observations of advanced stage mergers have found a range molecular gas depletion times that are typically shorter than in local spirals \citep{Saito2015,Saito2016,Bemis2019,Wilson2019}, and the relationship between $\Sigma_{SFR}$ and $\Sigma_{mol}$ is steeper than unity.  However, as a class it is critical to consider the diversity of merger stages when considering its gas and star-formation content \citep{sanders1996,Combes1994,Larson2016}. \cite{Espada2018} finds that for wide separation merging systems this relationship can be shallower than in local spirals, and the depletion time can be longer toward galaxy centers.

In the local Universe recent technical advances now make it possible to measure the star formation efficiency per free-fall time,
\begin{equation}
\epsilon_{ff} = \frac{\Sigma_{SFR}}{\Sigma_{gas}/t_{ff}},
\label{eq:eff}
\end{equation}  
at spatial scales of $\sim$100~pc  in nearby galaxies \citep{Leroy2015,Hirota2018,Utomo2018}. When isolating the star forming regions it is typically safe to assume that the gas mass surface density, $\Sigma_{gas}$, can be approximated by the molecular gas mass surface density, and therefore later in this work we will use $\Sigma_{mol}$ as an approximation of the star forming gas. The star formation efficiency per free-fall time takes the three-dimensional shape of the cloud through the estimation of the free-fall time, 
\begin{equation}
t_{ff}\equiv \sqrt{\frac{3\pi}{32G \rho}}.
\label{eq:tff}
\end{equation}
Where $\rho$ is the volume density of the region being measured. \cite{Utomo2018} measures $\epsilon_{ff}$ in local spirals at $\sim120$~pc resolution with a method that is similar to what we use.  They find a typical $\epsilon_{ff}\approx 0.5$\%. Detailed studies of nearby spiral galaxies M~51 \citep{Leroy2017} and M~83 \citep{Hirota2018} find low values, consistent with \cite{Utomo2018}. Yet, those studies also show there may be systematic variation in the values of $\epsilon_{ff}$ at the 0.3~dex level, suggesting a completely universal value may not be a correct assumption. 



The amount of variation of $\epsilon_{ff}$ both from galaxy-to-galaxy and within galaxies is important to star formation models. A number of theories make the explicit assumption that  star formation proceeds at a constant, ``low'' efficiency with $\epsilon_{ff}\approx1$\% \citep{krumholz2012,salim2015}. Other theories that do not explicitly assume this find very little variation in simulations \citep{shetty2012,kim2013}. If $\epsilon_{ff}$ varies significantly in different types of galaxies, this would limit the applicability to those theories.  Some theory and simulation predict that in very active regions, with very dense clouds, the efficiency can reach 10-30\% \citep{Murray2010,faucher2013,Grudic2019}.  


In this paper we present a map of CO(2-1) with $\sim$200~pc resolution in a starbursting face-on galaxy, IRAS08339+6517. The galaxy exhibits many properties in stellar populations, structure and kinematics that are similar to compact, turbulent disks more commonly found at $z\approx 1-2$. We measure the internal distribution of $t_{dep}$ and $\epsilon_{ff}$, as well as the $\sim$1~kpc resolved star-formation law, and consider these results in light of star formation models.

\begin{figure*}

\begin{center}
\includegraphics[width=\textwidth]{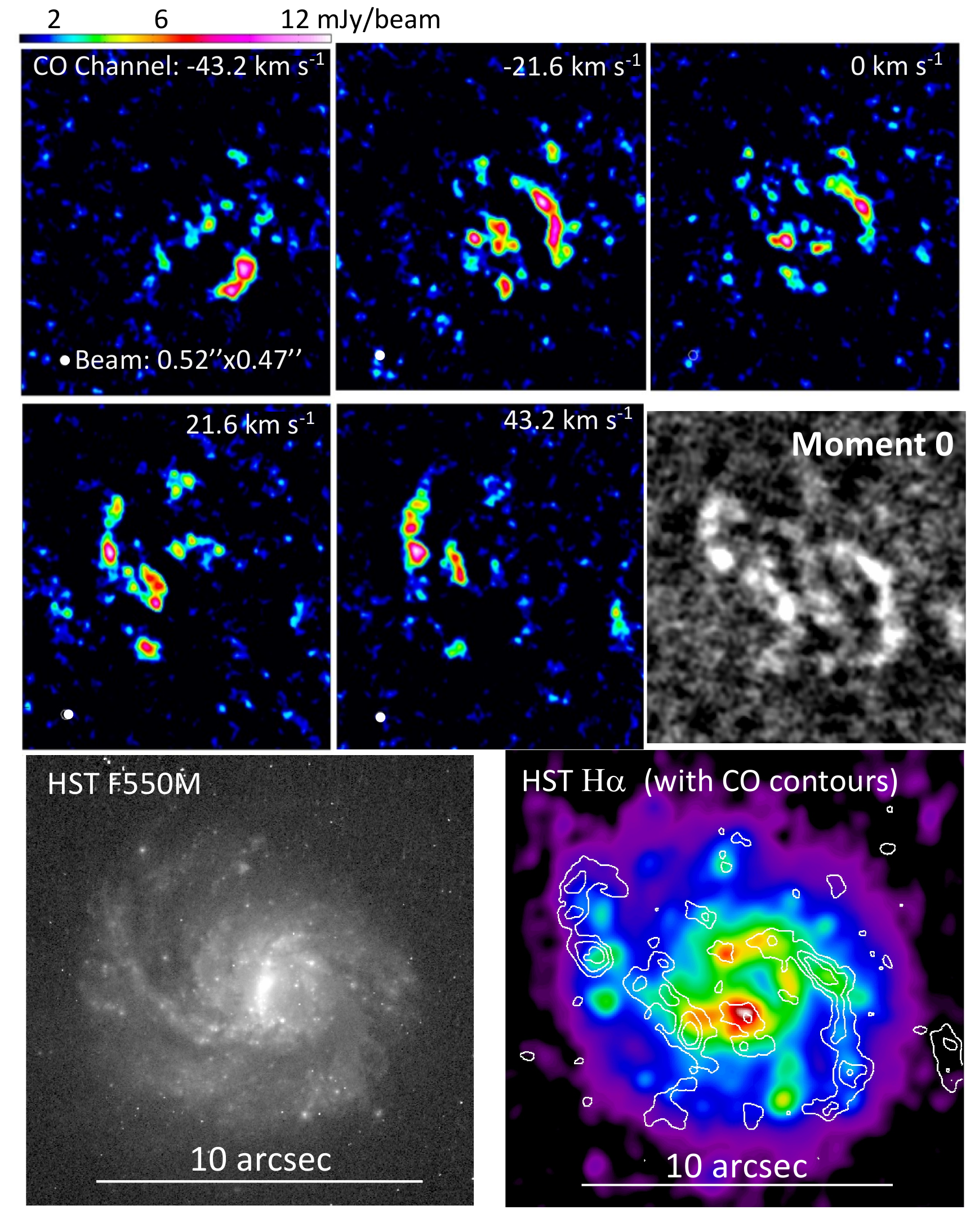}
\end{center}
\caption{The NOEMA CO(2-1) emission is shown in 5 velocity channels of width 21.6 km~s$^{-1}$. These channel widths are used for display purposes only and are chosen to isolate the clumps of CO gas. We span the velocity range of the galaxy.  In the bottom left panel we plot the star light (HST F550M). The white bar indicates 10 arcsec, which is roughly equivalenth to 4~kpc. 
In the bottom row, right panel we also overplot the CO(2-1) contours from a moment zero map onto the HST H$\alpha$ map convolved to matching spatial resolution as the CO data.  \label{fig:maps} }
\end{figure*}


\section{Methods}
\subsection{CO Observations and Molecular Gas Mass} 
We obtained CO(2-1) observations (Fig.~1) with the NOrthern Extended Millimeter Array (NOEMA). All observations use the new PolyFiX correlator tuned to sky frequency of 226.215~GHz in USB with a channel width of 2.7~km~s$^{-1}$ utilizing all 9~antennas. IRAS08 was observed for 13 hours in A~configuration on 18-Feb-2018, and on 01-Apr-2018 for 5.5~hr in C-configuration. By including the C-configuration data, and also considering the relative compactness of our source, we are likely not missing a significant amount of low-spatial frequency data, and are rather more strongly affected by point source sensitivity. The maximum recoverable scale of the C-configuration data is $\sim$6 arcseconds, which corresponds to $\sim$2.5~kpc. We can compare this to the half-light radius of the starlight, which is $\sim1$~kpc, or roughly 2.5 arcseconds. We should therefore recover twice the half-light radius of the stars.   

Observations were calibrated using {\em GILDAS} routines in {\em CLIC}, and then cleaned with the {\em MAPPING} pipeline routine during an on-site visit to IRAM. We achieve a point source sensitivity of 1.4~mJy~beam$^{-1}$ in 20~km~s$^{-1}$ of bandwidth, and beam size of $0.52\times0.47$~arcsec$^{2}$ ($\sim 197\times178$~pc$^2$).

In this paper we consider two scenarios for CO-to-H$_2$ conversion. First, we use the standard Milky Way conversion of $\alpha_{CO}=4.36$~M$_{\odot}$ (K~km~s$^{-1}$~pc$^{-2}$)$^{-1}$ and a line ratio of $R_{12}$=CO(2-1)/CO(1-0) = 0.7  ($\alpha_{CO}^{2-1} = \alpha_{CO}/R_{12}$).  The metallicity  (0.7~Z$_{\odot}$, \citealp{Lopez2006}) and morphology (Fig.~\ref{fig:maps}) both suggest a Milky Way like conversion factor. Second, the large IR flux ratio, $f_{60}/f_{100}\approx0.8$ \citep{Wilkind1989} suggests  a value of $\alpha_{CO}\approx1.8-2.5$~M$_{\odot}$ (K km~s$^{-1}$~pc$^{-2}$)$^{-1}$ \citep{magnelli2012xco}. 
We find a total molecular gas mass of 2.1$\times10^9$~M$_{\odot}$, using the Milky Way CO-H$_2$ conversion factor. This is similar to the total flux estimated from single dish observations by \cite{Wilkind1989}, and we are thus not likely missing significant amounts of flux. 

Using the kinematics from \cite{Cannon2004} observations of HI gas we can estimate the total mass of the system. There is, however, an added source of uncertainty in that the disk is relatively face-on, which makes estimating the circular velocity uncertain by the inclination angle. Moreover, the large radius HI gas is interacting with a companinion (described below) and is likely not a good indicator of the total mass. Nonetheless, we can determine if derived mass from CO strongly disagrees with this, as a sanity check on our measurement. The velocity of the HI does cleanly assymptote to a flat curve, but rather turns around at 15~kpc due to the interaction. It is not clear what appropriate $v_{circ}$ to assume. We opt for a value closer to the galaxy of $\sim100$~kpc, though we note that total mass depends strongly on $v_{circ}$, and even slightly larger values give significantly larger total masses. We assume an inclination angle of 20$^o$ based on the average of the F550M isophotes. Using the HI kinematics derived in \cite{Lopez2006} and a galaxy size of $2\times R_{1/2}\approx 2$~kpc, based on the star light. This gives a total mass of $\sim1.2\times10^{10}$~M$_{\odot}$. The total stellar mass is $\sim10^{10}$~M$_{\odot}$ (described below). \cite{Cannon2004} finds that the HI mass associated to the galaxy of IRAS08 is $\sim0.11\times10^{10}$~M$_{\odot}$. Therefore, our derived molecular gas mass of 2.1$\times10^9$~M$_{\odot}$ is roughly consistent with kinematic observations.

We note that even the bimodal assumption of $\alpha_{CO}$ as either Milky Way value or star burst may be an over-simplification for high $\Sigma_{mol}$ galaxies. These galaxies may possibly have an $\alpha_{CO}$ that varies with local mass surface density \citep{narayanan2011,bolatto2013}. In the text, we will consider the impact of this on our results. 

We measure the resolved properties of the CO(2-1) map using the moments of the data cube. These are measured on the 2.7~km~s$^{-1}$ spectral resolution data cube, and the spaxels are binned to 0.51~arcsec, matching the circularised FWHM. We use an interpolation intended to conserve flux when regridding. We check this by measuring the flux in an identical circular region 6 arcsec in radius. We find the flux is the same to 99\%. Using the CASA task {\em immoments} we determine the integrated intensity, velocity and velocity dispersion of the CO(2-1) line in each resolution element. To calculate the velocity moments we only include data with $S/N>3$, and in a region of the spectrum that is restricted to contain the emission line of the galaxy. 
\begin{figure*}
\begin{center}
\includegraphics[width=0.8\textwidth]{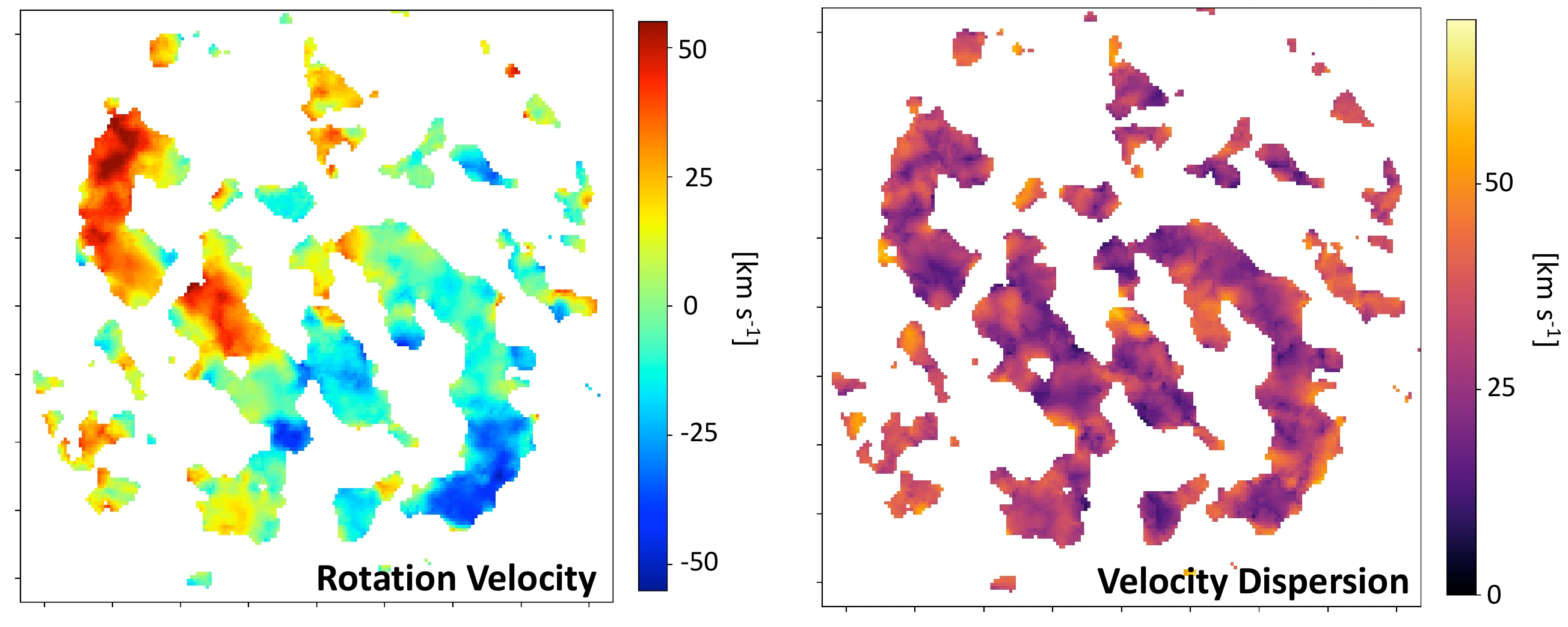}
\end{center}
\caption{We  plot the velocity of CO(2-1), showing rotation (left) and velocity dispersion (right) of IRAS08. The galaxy is rotating and as shown later the velocity dispersion is relatively flat with respect to position int he galaxy. We note that due to our sensitivity limits the observations are biased toward the bright CO emission line sources. We cannot say if the velocity dispersion is significantly different off of the arms. \label{fig:velmap}}
\end{figure*}

\subsection{Resolved Star Formation Rate}
To measure the star formation (SFR) in IRAS08 we use H$\alpha$ image produced from {\em Hubble Space Telescope} observations \citep{ostlin2009}, convolved to the beam of our CO observations. Continuum subtraction was performed by modelling the stellar continuum with multi-band photometry \citep{Hayes2009}.  The measurement uncertainty of H$\alpha$ flux for individual clumps is less than 1\% in all cases. 

We determine the extinction by stellar population fits using the CYGALE \citep{Boquien2018arxiv} fitting code to HST/ACS image filters SBC F140LP , HRC F220W, HRC F330W, WFC F435W, and WFC F550M. 
We measure, and correct for, the extinction in individual regions that are set to match the resolution of the CO(2-1) map. Averaged over the whole galaxy our fits recover A$_{H\alpha}\approx0.2$~mag, which similar to previous results using line ratios \citep{Leitherer2002,Lopez2006,ostlin2009}. To correct the H$\alpha$+[NII] narrow-band images to H$\alpha$, we use the median [NII]/H$\alpha$ ratio from the longslit data in \cite{Lopez2006}, which is [NII]/H$\alpha\approx 10$\%. This is consistent with expectations for a moderately low metallicity galaxy \citep{Kewley2019Review}.

To convert the ionized gas flux to SFR we use the calibration \citep{hao2011} $SFR = 5.53\times 10^{-42}L_{H\alpha}$, where $L_{H\alpha}$ is the extinction corrected luminosity of H$\alpha$ gas in units of ergs~s$^{-1}$, which assumes a Kroupa IMF.

A particular concern in deriving resolved star formation rates of both LIRG and UV bright galaxies is the possible presence of AGNs in the central parts of the galaxy. There have been a number of works that have analysed the optical and UV spectra of IRAS08, which we can use to motivate our interpretation of the H$\alpha$ flux as coming from star formation. \cite{otifloranes2014} carries out extensive modelling of the X-ray, UV and optical data from the center of IRAS08, and finds it consistent with a super-star cluster with age 4-5~Myr. Similar, \cite{Lopez2006} does not find elevated, nonthermal line ratios in the galaxy center with respect to the rest of the galaxy. Similarly, in out KCWI data (described below) we do not see a significant change in line ratios, for example [OIII]/H$\beta$, in the galaxy center that would suggest the driving mechanism of the emission line is changing. We find that [OIII]/H$\beta$ in the galaxy center is similar to that of the outer disk, at the 0.1-0.2~dex level. Future observations that can compare directly [OIII]/H$\beta$ to [NII]/H$\alpha$ would be definitive, see \cite{Kewley2019Review} for recent review. Nonetheless, at present there is not any evidence to suggest a prominent AGN in the center of IRAS08. 
 
\subsection{Metallicity Measurement}
In this paper we will use the metallicity profile as a signature of possible gas inflows \citep{Kewley2010}. To calculate the metallicity in IRAS08 we use [OII], [OIII] and H$\beta$ observations take from the Keck Cosmic Web Imager \citep{Morrissey2018}. The galaxy was observed for 20 minutes, using the BM grating in Large Field mode with two central wavelength settings of 405 \& 480~nm. Data was reduced with standard KCWI pipeline methods\footnote{https://github.com/Keck-DataReductionPipelines/KCWI\_DRP/} using in-frame sky subtraction. Before the sky subtraction step, we masked out all galaxy continuum and emission features for accurate sky estimates.  Field-of-view of the Large slicer is 33''$\times$20.4''. KCWI is seeing limited and spaxels have a dimensions of 0.7$\times$1.35~arcsec$^2$ (279$\times$600~pc$^2$). 

The metallicity was derived in each spaxel using the so-called R23 method, as described in \cite{kobulnicky2004}. 
The metallicity of IRAS08 is near the branching point of the R23-metallicity calibration. We therefore take an iterative approach to solving for metallicity. We start by assuming that each spaxel is on the ``upper branch" ($12+log(O/H)>8.5$) and solve for the metallicity. If the derived metallicity is less than 8.5, we then recalculate that spaxel on the lower branch. This is carried on for a 3 iterations. We then re-do the procedure starting on the lower branch. We find that the absolute value of the metallicity changes by $\pm 0.4$~dex depending on the branch choice, but the gradient of the metallicity across the disk is the same whether we start on the upper or lower branch. We restrict our analysis to the gradient of metallicity, and only make direct, quantitative comparison to other measurements using the same emission lines, as different metallicity indicators may yield different gradients. 

\section{Properties of IRAS08339+6517}
IRAS08339+6517 (hereafter IRAS08) is a face-on galaxy with redshift of $z\approx0.0191$, which translates to a luminosity distance of $\sim$83~Mpc. The galaxy is known to be UV-bright, compact, and have young stellar populations \citep{Leitherer2002,Lopez2006,overzier2008}. The global mass-weighted age of IRAS08 is quite young compared to local spirals with published age estimates varying between 10-50~Myr \citep{Leitherer2002,Lopez2006}. In this section we discuss the resolved properties of the gas morphology and kinematic state of IRAS08, with  emphasis on how IRAS08 is an outlier for local galaxies, and often is more similar to galaxies at $z\approx 1.5$. We also discuss the interaction that IRAS08 is experiencing with a distant, smaller companion galaxy.  

The basic properties described in this section are summarized in Table~\ref{table:iras08}. 



\subsection{Total Mass, Size and SFR}
Using the optical colors from \cite{Lopez2006}, and assuming a Kroupa IMF, we estimate a K-band mass-to-light ratio of $log(M/L_K)\approx$0.3-0.4, depending on the model assumptions \citep{bell2003,Zibetti2009}. Using the K-band magnitude of 11.88~mag and log($M/L_K$)$\approx$0.35, the total stellar mass of IRAS08 to be $M_{star}\approx$1.1$\pm0.3\times10^{10}$~M$_{sun}$.  

The size of the galaxy is determined by measuring the surface photometry of the F550M ACS/WFC image, which is roughly V-band. We use the same software and technique as developed in \cite{fisherdrory2008}. We find that the half-light radius of IRAS08 is 2.54 arcseconds, which translates almost exactly to 1.0~kpc. This makes IRAS08 a $\sim$2-3$\sigma$ outlier toward smaller sizes (more compact) than what is expected from the $r_e-M_*$ relationship measured on $z\approx0$ galaxies \citep{Mosleh2013}. We can also compare the size of IRAS08 to local Universe galaxies of similar IR brightness. \cite{Arribas2012} finds that LIRGS in general have a median H$\alpha$ half-light radius of $\sim$2~kpc. They show that the largest LIRGs are pre-coalescence systems, like IRAS08, the median H$\alpha$ half-light radius of pre-coalescence LIRGs is closer to $\sim 3$~kpc. IRAS08 is thus more compact than the median LIRG, especially those that are pre-coalescence.  

\begin{deluxetable}{lc}
\tablewidth{0pt} \tablecaption{Basic Properties of IRAS08339+6517}
\startdata
\hline \\
Total Stellar Mass & 1.1$\pm0.3\times10^{10}$~M$_{\odot}$\\
Total SFR  & 12.1$\pm1$~M$_{\odot}$~yr$^{-1}$\\ 
Stellar $R_{1/2}$ & 1~kpc \\
Molecular Gas Mass$^a$ & 2.1$\times 10^9$~M$_{\odot}$\\
HI Gas Mass & 1.1$\times$10$^9$~M$_{\odot}$\\
Molecular gas velocity dispersion & 25$\pm6$~km~s$^{-1}$\\
Molecular gas depletion time & 0.12~Gyr \\
SFR/SFR$_{MS}$ & 12$\times$ \\ 
Toomre $Q_{gas}$ & 0.5 \\
\\
\enddata
\label{table:iras08} 
\tablenotetext{a}{$\alpha_{CO}=4.36$ M$_{\odot}$ (K km s$^{-1}$ pc$^{-2}$)$^{-1}$ }
\end{deluxetable}

We estimate the total SFR using the integrated H$\alpha$ luminosity of 1.8$\times10^{42}$~erg~s$^{-1}$ and $A_{H\alpha}=0.19$ \citep{ostlin2009}. Using the calibration from \cite{hao2011}, assuming a Kroupa IMF, we find SFR=12.1~M$_{\odot}$~yr$^{-1}$. \cite{sanders2003} finds a 25~$\mu$m flux of 1.13~Jy. We can thus also estimate the total SFR from the combined H$\alpha$ and 25~$\mu$m fluxes using the calibration from \cite{kennicutt2009}, which gives a very similar value of 11.1~M$_{\odot}$~yr$^{-1}$. For consistency with resolved SFR measurements we will use the H$\alpha$ only value for calculations. The SFR for IRAS08 is 12 times the value of the $z=0$ main-sequence for a similar mass galaxy \citep{Popesso2019}. 

The global, galaxy-averaged molecular gas depletion time of $t_{dep}\equiv\Sigma_{gas}/\Sigma_{SFR}\approx 0.12$~Gyr. \cite{tacconi2017} find a relationship between offset from the main-sequence such that $t_{dep}$ $t_{dep}(\delta MS) \approx (1+z)^{-0.6}\times \delta MS^{-0.44}$. For $\delta MS\approx 12$ and $z=0$, this translates to a value of 0.33~Gyr. Interestingly, in spite of its many idiosyncrasitic properties, IRAS08 is behaving similar to other star-bursting galaxies (in terms of $t_{dep}$). 

The molecular gas fraction for IRAS08 is $f_{gas}\equiv M_{mol}/(M_{star}+M_{mol})\approx0.17$. This gas fraction is in the top 95$^{th}$ percentile of gas rich galaxies in the local Universe, using COLD GASS survey as a $z\approx 0$ benchmark \citep{saintonge2011a}.

\subsubsection{Large Gas Velocity Dispersion \& Low Toomre Q}
Here we discuss the internal kinematic state of IRAS08. In IRAS08 we find through direct, resolved measurement that the disk is consistent with being marginally stable to unstable ($Q\approx0.5-1.0$), and has elevated molecular gas velocity dispersion, compared to local spirals, across all radii. 

\begin{figure}
\begin{center}
\includegraphics[width=0.49\textwidth]{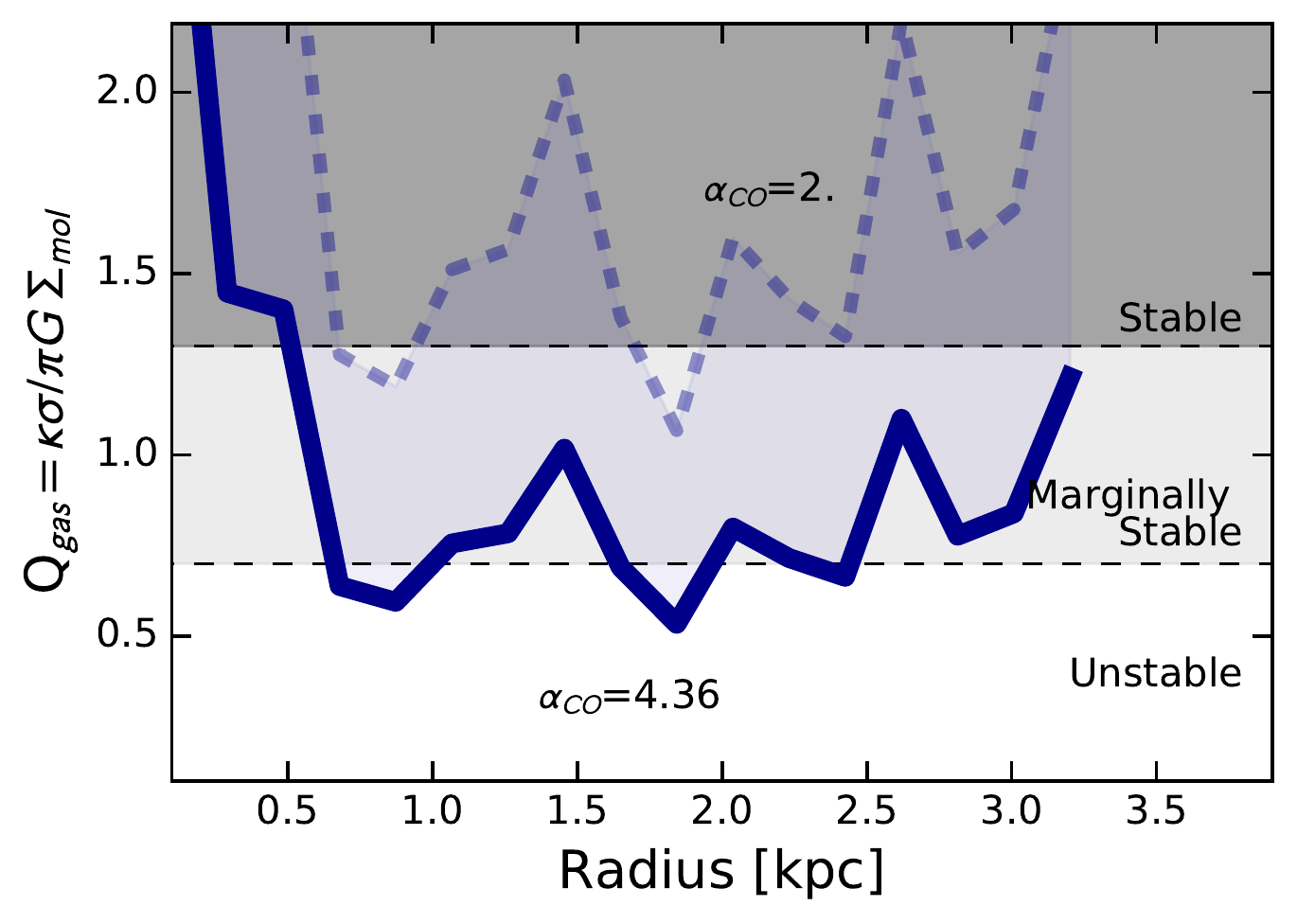}
\end{center}
\caption{ The radial profile of Toomre Q for the molecular gas disk in IRAS08 is shown. The solid line represents the profile for the Milky Way $\alpha_{CO}$, and the dashed blue line for the star burst conversion factor.  The Q-profile suggests that IRAS08 is most likely consistent with an unstable disk ($Q\approx0.5-1.3$) for almost all radii across the molecular disk. Note that the upper bound for stable versus marginally stable is arbitrarily defined to guide the eye. We set this line at $Q\approx1.3$ based on typical systematic and measurement uncertainties in the value of Q \citep[e.g.][]{genzel2011}.
\label{fig:qprof}}

\begin{center}
\includegraphics[width=0.5\textwidth]{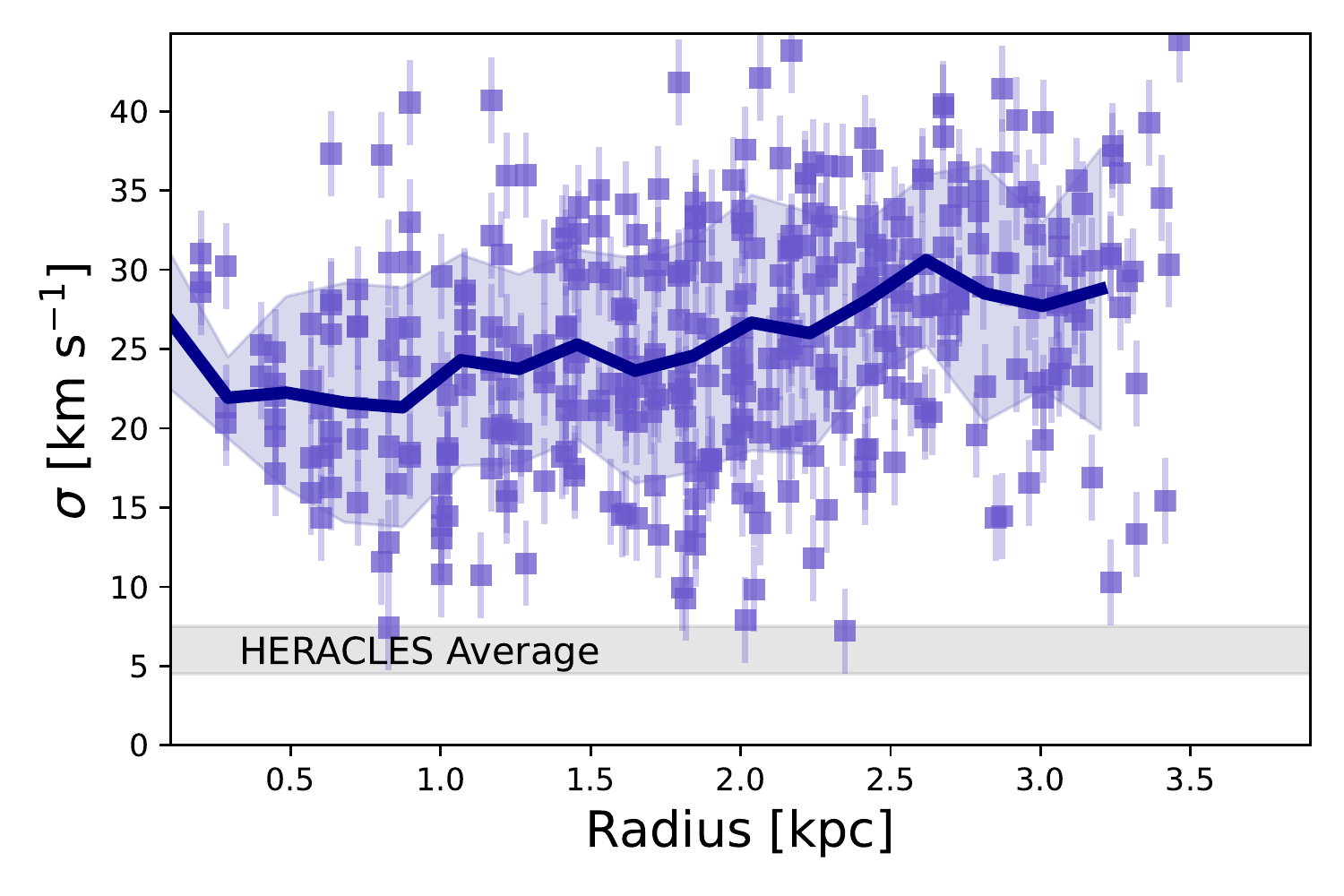}
\end{center}
\caption{The velocity dispersion, $\sigma$, is plotted as a function of galactocentric radius for IRAS08 for all lines of sight within the area of the galaxy (squares), as well as the average (dark line) within 0.5 arcsec ($\sim$200~pc) bins. The shaded region represents the standard deviation about the running average. The measured $\sigma$  in IRAS08 is significantly larger than is typical for local spirals, which is indicated by the grey horizontal bar \citep{Ianjamasimanana2012} , and only mildly changes  with radius.
\label{fig:sig_prof}}

\end{figure}

The stability of a self-gravitating disk is characterised by Toomre's $Q$ \citep{toomre1964,binneyandtremaine}, where 
\begin{equation}
Q_{gas} \equiv \frac{\kappa \sigma_{gas}}{\pi G \Sigma_{gas}}.
\label{eq:q}
\end{equation}
The quantity $\kappa$ is the epicyclic frequency. We measure it directly as $\kappa^2=4(v/R)^2+R d(v/R)^2/dR$. We determine the velocity, $v$, by fitting the flat disk model, $v(R) = v_{flat}[1-exp(-R/r_{flat})]$, to the velocity map, shown in Fig.~\ref{fig:velmap}. We assume an inclination of 20$^o$ based on the average ellipticity of the isophotes in F550M image. We note that for low inclination galaxies the rotation field is particularly uncertain, and should not be used on its own be a deciding factor in determining the physical state of the galaxy. We take the $Q$ value only as one aspect of the IRAS08.  The velocity dispersion, $\sigma$, is measured from the CO moment map. An infinitely thin disk is considered unstable if $Q<1$. Disks of non-zero thickness, like those with high velocity dispersion, are unstable if $Q\lesssim 0.7$ \citep{Romeo2010}.  

In IRAS08 we find that the median $Q\approx0.5$, assuming the Milky Way $\alpha_{CO}$. In Fig.~\ref{fig:qprof} we show that $Q(R)$ remains marginally unstable across all radii except the very center of the galaxy, where $\kappa$ becomes very large. Though $Q\approx0.5$ may seem extreme by local Universe standards, we reiterate that for galaxies with a thicker disk, as indicated by high gas velocity dispersion, the critical value of stability using Equation~$\ref{eq:q}$ is 0.7. Moreover, this value of $Q$ is similar to the values of $Q$ in the DYNAMO galaxies \citep{fisher2017apjl}, which have similarly high SFR and gas content. Moreover, it is essentially the same calculation using CO kinematics and surface density as observations of a CO disk at both $z\approx1.5$ \citep{genzel2013} and $z\approx4$ \citep{tadaki2018}, and find a very similar value of $Q_{gas}\approx 0.5-1.0$. Conversely, Toomre-$Q$ measured on local spirals produces $Q_{gas}\approx 2-10$ \citep{leroy2008} over a wide-range in radius. For IRAS08, the assumption of lower $\alpha_{CO}$, motivated by the high dust temperature, still keeps the galaxy in the ``stable" regime. It remains lower than $Q_{gas}$ values seen in THINGS survey spiral galaxies.  

In Fig.~\ref{fig:sig_prof} we show that the galaxy averaged velocity dispersion in IRAS08 is $\sigma \approx 25$~km~s$^{-1}$ with a root-mean-squared deviation (RMS) of $\pm6$~km~s$^{-1}$. The velocity dispersion is taken from the moment map, with the channel width removed in quadrature. This however has a very small effect as the channels are less than 5\% of the typical FWHM of the line. Because the galaxy is face-on, well resolved ($\sim$200~pc), and has a relatively low rotation velocity the velocity gradient across individual spaxels due to rotation is small. Using the velocity model fit to the moment~1 map, we find that in the central resolution element the line dispersion introduced by velocity gradients is $\sigma_{vel}\sim6$~km~s$^{-1}$. This is calculated by taking the model fit to the velocity field and resampling it to match our beam size. The velocity field begins with an assumption of infinitely thin emission lines, and when sampled at the beam resolution the width reflects the range of velocities in the beam.  Removing this in quadrature would alter the measured $\sigma$ by less than 1~km~s$^{-1}$. We, therefore, do not make a correction for beam smearing as it is a small effect, and likely introduces its own systematic uncertainties. 

\cite{Ianjamasimanana2012} find that the average CO velocity dispersion of galaxies in the HERACLES sample is 6~km~s$^{-1}$ with a standard deviation of 1.5~km~s$^{-1}$. The velocity dispersion of IRAS08 is 4$\times$ what is measured in local spirals, making it a $\sim$10$\sigma$ outlier. Work using stacking of CO spectra uncovers a secondary broad component of dispersion with $12\pm4$~km~s$^{-1}$ \citep{Caldu2013}. The unstacked velocity dispersion of IRAS08 remains a 3$\sigma$ outlier from the broad component in local spirals.

It is difficult to compare the velocity dispersions to that of LIRGS, as there are not many studies of the resolved gas velocity disperion, especially targeting molecular gas in LIRGS. Also, LIRGs represent a very diverse set of objects when considering morpho-kinematic properties \citep[e.g.][]{Larson2016}, and it is, thus, challenging to make a well-posed comparison. \cite{Espada2018} makes similar resolution maps of two LIRGs with wide-seperation interactions. They found that $\sigma_{mol}$ varies much more in their targets than we find in IRAS08. In their targets the disk has a low dispersion, with $\sigma_{mol}\sim 10-20$~km~s$^{-1}$ and the center is higher, $\sim$40~km~s$^{-1}$. \cite{Zaragoza2015} studies resolved kinematics in samples of interacting, though not advanced, merging galaxies. They find the intercting systems have a median $\sigma\sim10-15$~km~s$^{-1}$. There is, therefore, a range of velocity dispersions in local Universe LIRGs galaxies and it is difficult to make any conclusive statement about the comparison. 



\subsection{Distant Interaction}
There is a plume of HI gas that extends from IRAS08 in the direction of a nearby companion \citep{Cannon2004}.  The stellar light of the companion galaxy is quite low compared to IRAS08($\sim$1/10-1/20; \citealp{Lopez2006}) and the separation is $\sim$60~kpc. For this separation and mass ratio the merger classifications developed on GOALS sample galaxies \citep{Larson2016} places IRAS08 in the minor-merger category. 
We do not observe signs that the interaction in IRAS08 \citep{Cannon2004,Lopez2006} is directly altering either the morphology or kinematics of IRAS08 inside the 90\% radius of the F550M image.  (1)  The morphology of starlight (Fig.~1) does not, show signs of significant disturbances (such as in advanced mergers like Antennae galaxies). There is an asymmetry to the spiral arms. We measure the asymmetry value of 0.17-0.2 depending on whether we use the 50\% or 90\% radius, respectively. \cite{Conselice2014} reviews galaxy morphology and finds asymmetries of 0.15$\pm$0.06 for typical late-type disks, and 0.32$\pm$0.19 for ULIRGs. IRAS08 falls in between the two values.  It is on the high side, but within the distribution of late-type disks, and on the low-side of ULIRGS. It is below the range of asymmetries that are quoted as ``typical" for starbursts.  (2) The stellar light profile is well described as a smooth exponential decay \citep{Lopez2006}. (3) The kinematics inside the radius of the galaxy are well-fit with a rotating disk model (Fig.~1). (4) There is not a significant off-center rise in the velocity dispersion. These are similar criteria as used in studies of galaxies at $z\approx 1-3$ to classify mergers and rotating galaxies \citep[e.g.][]{forsterschreiber2009,genzel2011}. Based on these observations, it does not appear appropriate to categorize IRAS08 with advanced stage mergers. The main impact of this interaction could be that it provides the disk of IRAS08 with a supply of gas that is of order the gas mass inside the disk $\sim 4\times10^9$ M$_{\odot}$ \citep{Cannon2004}. We note that the plume HI gas could be an outflow from the star burst of IRAS08, however, this would open the question as to why there is not a symmetric flow on the opposite side, as expected from biconical flows. In the Appendix we consider the interaction as possible driver of the $t_{dep}$ properties of IRAS08, though we find it behaves differently than other galaxies with similar interaction parameters. 

\begin{figure*}[h!]
\begin{center}
\includegraphics[width=0.95\textwidth]{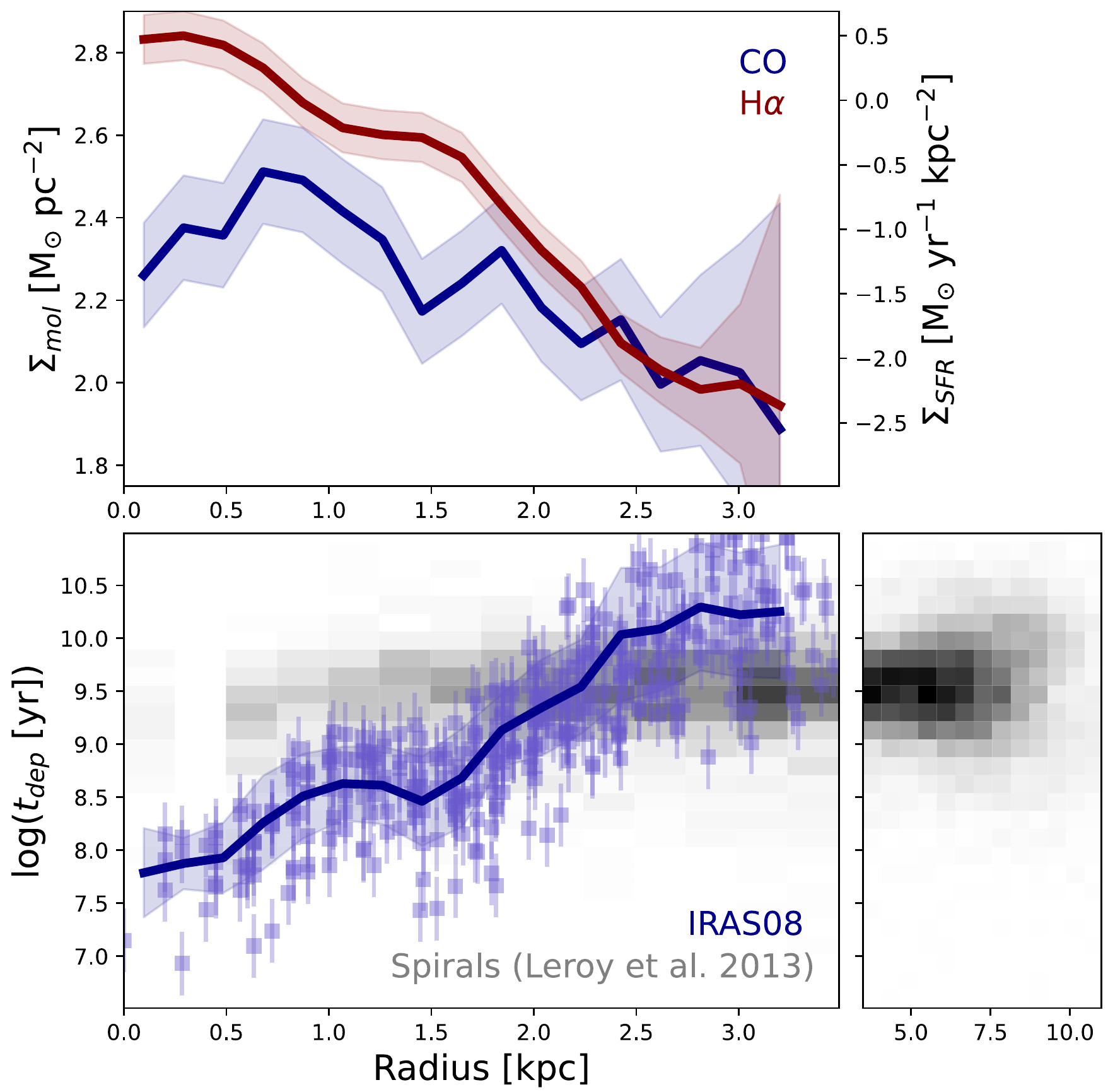}
\end{center}
\caption{The radial profile of the molecular gas mass surface density (top), SFR surface density (top) and molecular gas depletion time (bottom) in IRAS08 is shown. For the depletion time measurements of individual beams are shown as the blue squares. In both panels, the radial average in 200~pc increments is shown as a solid line, with the standard deviation shown as a shaded region. The uncertainty on $t_{dep}$ due to $\alpha_{CO}$ is shown as the errorbar. For comparison, we also show data from the HERACLES survey \citep{leroy2013} for the central 4~kpc  of galaxies, and in a compressed panel to the right we show $R_{gal}=5-10$~kpc for HERACLES galaxies. We display the HERACLES data in this manner to emphasize the extreme nature of IRAS08, not only is the gradient strong compared to typical spirals, but this occurs over a very compact radius, despite have a similar total mass.  In IRAS08 $t_{dep}$ is a strong function of radius, compared to local spirals which are essentially constant over  10~kpc in radius. In the top panel we show that the decrease in $t_{dep}$ is does not monotonically correlate with an increase $\Sigma_{mol}$. 
\label{fig:tdep_prof}}
\end{figure*}

\section{Spatial Variation in molecular gas depletion time} 
In this section we investigate the spatial variation of $t_{dep}$. We find that in IRAS08 $t_{dep}$ is of order $\sim$1.5-2~dex lower in the galaxy center than the outer parts of the disk. In large samples of spiral galaxies the most extreme variation observed is only of order $\sim$0.5~dex \citep{Utomo2017,leroy2013}.

\vskip 8pt
In Fig.~\ref{fig:tdep_prof} we show that there is a clear gradient in $t_{dep}$ within IRAS08 that is mush stronger than what is observed in HERACLES disk galaxies \citep{leroy2013}. 
We find that $t_{dep}$ increases from less than 0.1~Gyr in the central kiloparsec to greater than 3~Gyr at radii beyond the 80\% radius of the star light ($\sim$2.5~kpc). We measure a total gradient in the depletion time of $\Delta t_{dep}/\Delta R\approx7$~Gyr~kpc$^{-1}$. The local increase in $t_{dep}$ at $R_{gal}\sim1$~kpc by 0.2~dex is associated with an H$\alpha$ ring that contains high surface densities of CO. In Fig.~\ref{fig:tdep_prof} spirals are represented by HERACLES galaxies \citep{leroy2013}.  We measure a gradient in $t_{dep}$ of HERACLES galaxies from 0 to 3~kpc of $\Delta t_{dep}/\Delta R\approx0.3$~Gyr~kpc$^{-1}$. From $\sim$1-10~kpc the gradient is consistent with $\sim$0.  

We find in IRAS08 that $t_{dep}$ is more strongly coupled to $\Sigma_{SFR}$ than to $\Sigma_{mol}$. In the top panel of Fig.~\ref{fig:tdep_prof} we show the radially averaged profiles of $\Sigma_{SFR}$ and $\Sigma_{mol}$, the constituent components of $t_{dep}$. The profiles are plotted such that they overlap at large radius. The profile for $\Sigma_{SFR}$, is to low order approximation decreasing with radius at all points to the edge of the disk. Conversely, molecular gas shows a peak at $\sim$0.8~kpc and decrease in the galaxy center. We measure a correlation coefficient and $p$-value for both $t_{dep}-\Sigma_{SFR}$ and $t_{dep}-\Sigma_{mol}$. Of course both are strong correlations as they are circularly dependent, however we find that $\Sigma_{SFR}$ has a stronger correlation with $r=-0.98$ and $p\approx10^{-11}$, where molecular gas surface density has $r=-0.79$ and $p\approx10^{-5}$.  

The mostly likely impact of systematic uncertainties in the measurement of $\Sigma_{mol}$ would steepen the gradient of $t_{dep}$ for IRAS08. \cite{bolatto2013} argues that $\alpha_{CO}\propto \Sigma_{gas}^{-2}$ for values above $\sim$100~M$_{\odot}$~pc$^{-2}$ \citep[also see][]{sandstrom2013,narayanan2011}. This would lead to increasingly lower values of $\Sigma_{mol}$ than what is shown in Fig.~\ref{fig:tdep_prof} at radii $\lesssim$2.5~kpc, and thus even lower values of $t_{dep}$. We can also consider missing  low-spatial frequency emission from interferometic data. Though our NOEMA C-configuration observations likely assuage this, there could be an extremely flat distribution of CO gas that is filtered out by the interferometric observations. One would expect the low surface brightness extended emission to become more prominent at lower $\Sigma_{mol}$, i.e. larger radius, and thus this effect would also likely steepen the gradient of $t_{dep}$ in IRAS08. The decrease in $t_{dep}$ in the very center of the galaxy, from the ring at $R_{gal}\sim0.5$~kpc inward, could be due to the presence of an AGN increasing the H$\alpha$ flux. However, as we have discussed in the \S~2.2, there is not strong evidence for non-thermal emission from the currently observed emission lines. Moreover, the rise in H$\alpha$ flux in Fig.~\ref{fig:tdep_prof} covers an area that is larger than a single resolution element. Future, spatially resolved observations of the [NII]/H$\alpha$ ratio, combined with our current KCWI data would allow us to definitively determine what is driving the emission in each resolution element of IRAS08. Also, in comparison to the global trend in $t_{dep}$, the central part of the gradient is only a very small change. Overall, our estimates of the gradient in $t_{dep}$ err toward conservative estimates with respect to the systematic uncertainties.


We find that degrading the resolution acts to soften the gradient of $t_{dep}$ in IRAS08. We convolved the IRAS08 data cube to $\sim$800~pc resolution (4 beams). This resolution is chosen to be more similar to the resolution of the HERACLES data, while still being sufficient to sample the small size of IRAS08.  We find across the same radial range $\Delta t_{dep}/\Delta R\approx2$~Gyr~kpc$^{-1}$. This is still nearly an order-of-magnitude steeper than what is observed in HERACLES disks, and the central 800~pc has $t_{dep}$ in the range 60-140~Myr. \cite{Utomo2017} shows with a sample of spiral galaxies in the CARMA-EDGE that similarly the most common scenario is that $t_{dep}$ profiles are flat. They measure the ratio of $t_{dep}$ inside 1~kpc to the average of the rest of the disk, $t_{dep}^{cen}/t_{dep}^{disk}$. The most common value is unity. The most extreme targets have  $t_{dep}^{cen}/t_{dep}^{disk}\approx 0.1$. If we measure the same quantity in IRAS08 we find $t_{dep}^{cen}/t_{dep}^{disk}\approx 0.008$, implying a much steeper decline toward the galaxy center.


The radial gradient of $t_{dep}$ in IRAS08 is therefore in the range of 7-20$\times$ larger than in a typical local Universe disk galaxy from the HERACLES survey, and reaches depletion times that are $\sim$30$\times$ lower in the galaxy center than in the region surrounding the 90\% radius of the optical light.



\section{Star Formation efficiency per free fall time in IRAS08}
\subsection{Estimating free-fall time}

In IRAS08 we find values of $t_{ff}$ vary across the disk. The central values are $\sim$3~Myr and largest are $\sim$12~Myr. The central values skew to lower values than observed in local spirals \citep{Utomo2018}, but outer parts are similar. We discuss below that the systematic uncertainties here are of order 0.2~dex, and is enough to affect the comparison with local spirals. 

\vskip 5pt
As defined in Equation~\ref{eq:tff}, $t_{ff}\propto \rho^{-1/2}$. Measuring the volume density at or near the scale of clouds introduces a significant source of uncertainty. 
We follow the common approach to let $\rho = \Sigma_{gas}/(2 h_z)$, where $h_z$ is the scale height. If the gravitational potential is balanced by the kinetic energy then one can estimate that $h_z \propto \sigma_z^2/\Sigma$, and thus $\rho \propto (\Sigma/\sigma_z)^2$. Since IRAS08 is near to face-on, we can safely assume that $\sigma_z\approx \sigma$. 

One must, however, account for all sources of pressure support, such as magnetic fields and cosmic rays. Multiple prescriptions for this exist in the literature. To generate an estimate of the systematic uncertainty, we consider three recent calculations of the scale-height, and hence density. That of \cite{krumholz2017}, their Equation~22, such that $\rho \approx 2.8 G (\Sigma_{g}/\sigma)^2$, and that of \cite{Wilson2019}, their equation 7.  Thirdly, we  also consider the simple spherical-cloud assumption in which the galaxy has a constant disk thickness, and in each beam the gas is a sphere of radius $R_{cloud}\approx FWHM/2\approx 100$~pc, similar to other studies of star formation efficiency \citep[e.g.][]{Utomo2018}. 



The constant thickness assumption yields an average $<t_{ff}>= 6.47$~Myr with a standard deviation of 1.2~Myr. This is a factor of 2$\times$ shorter than the free-fall times measured in a similar fashion on local spirals \citep{Utomo2018}. The prescription used in \cite{Wilson2019} yields similar values of $t_{ff}$, with the constant thickness model on average shorter by a factor of 1.5-2$\times$.  The method of \cite{krumholz2017} gives larger values of $t_{ff}$, that are typically 3$\times$ what we estimate with the constant thickness  model. 
\begin{figure}
\begin{center}
\includegraphics[width=0.5\textwidth]{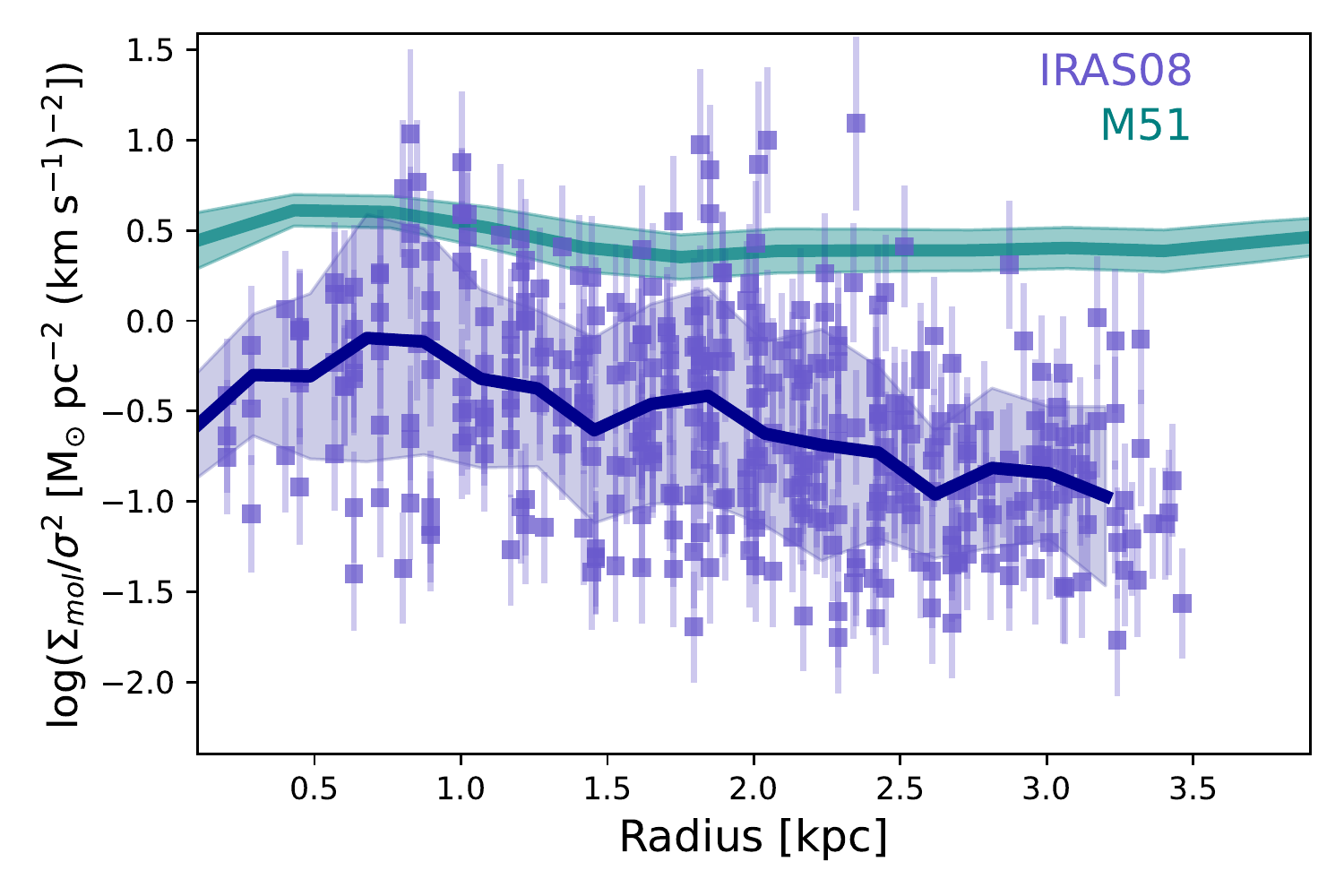}
\end{center}
\caption{ The quantity $\Sigma_{mol}/\sigma^2$ is plotted against radius for IRAS08 (blue) and M~51 (teal). The M~51 data is radial averages using data from \cite{Leroy2017}. These observables are assumed to correlate with the inverse of the disk scale-height. The lower values of $\Sigma_{mol}/\sigma^2$  in IRAS08 imply that, under the same assumptions, the IRAS08 disk is thicker than that of M51 by nearly an order-of-magnitude. The gradient in IRAS08 implies also that the disk scale height does not vary by more than $\pm$0.3~dex across IRAS08. \label{fig:b200} }
\end{figure}
\begin{figure}
\begin{center}
\includegraphics[width=0.5\textwidth]{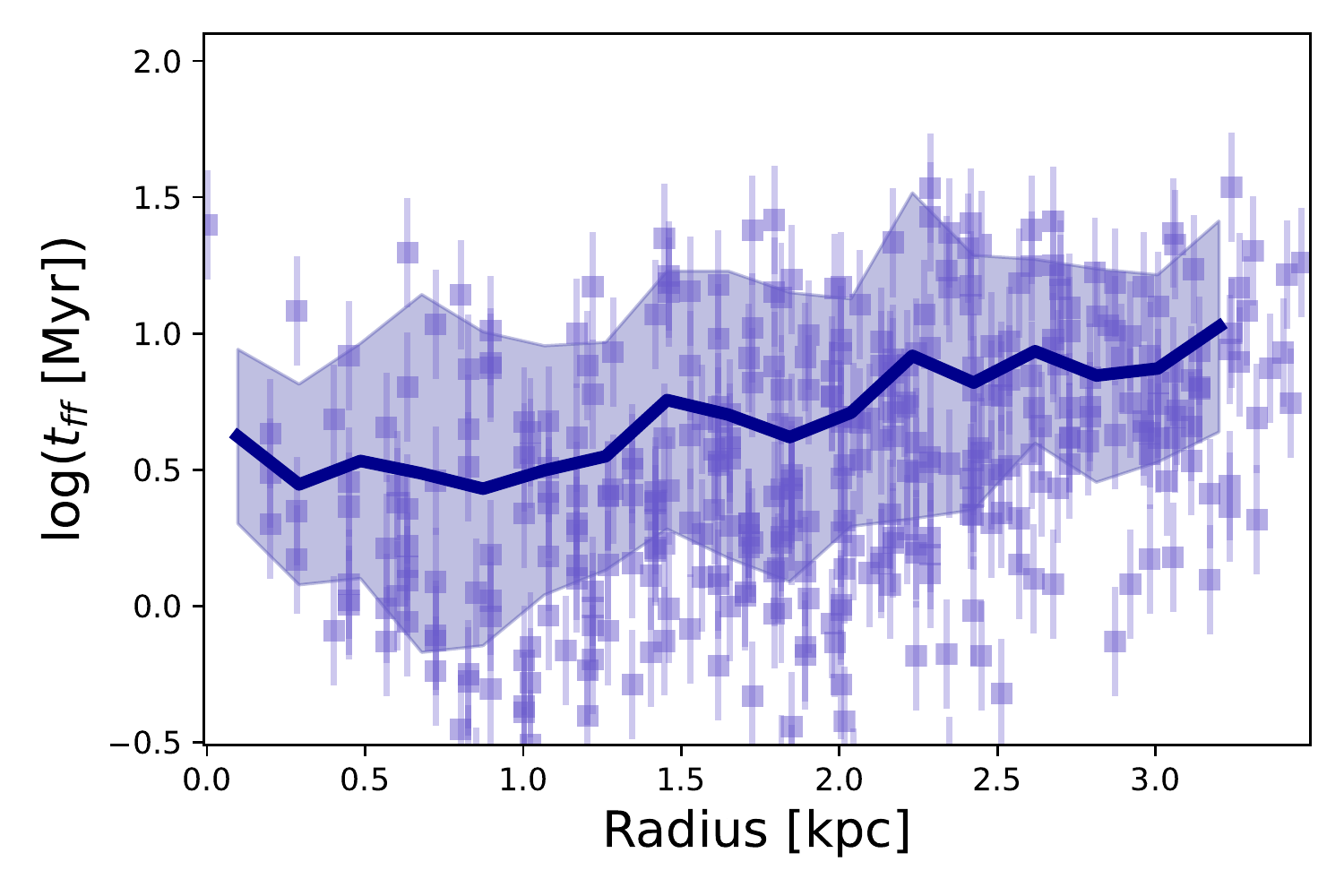}
\end{center}
\caption{ The free-fall time of all beams is shown for IRAS08. Here we plot $t_{ff}$ using the same prescription as in \cite{Wilson2019} for the disk scale-height. We find that across the disk the free-fall time varies by a factor of $\sim$3.5$\times$ from the center to the outer radii. This is not sufficient to account for the large change in $\epsilon_{ff}$.  \label{fig:tff} }
\begin{center}
\includegraphics[width=0.5\textwidth]{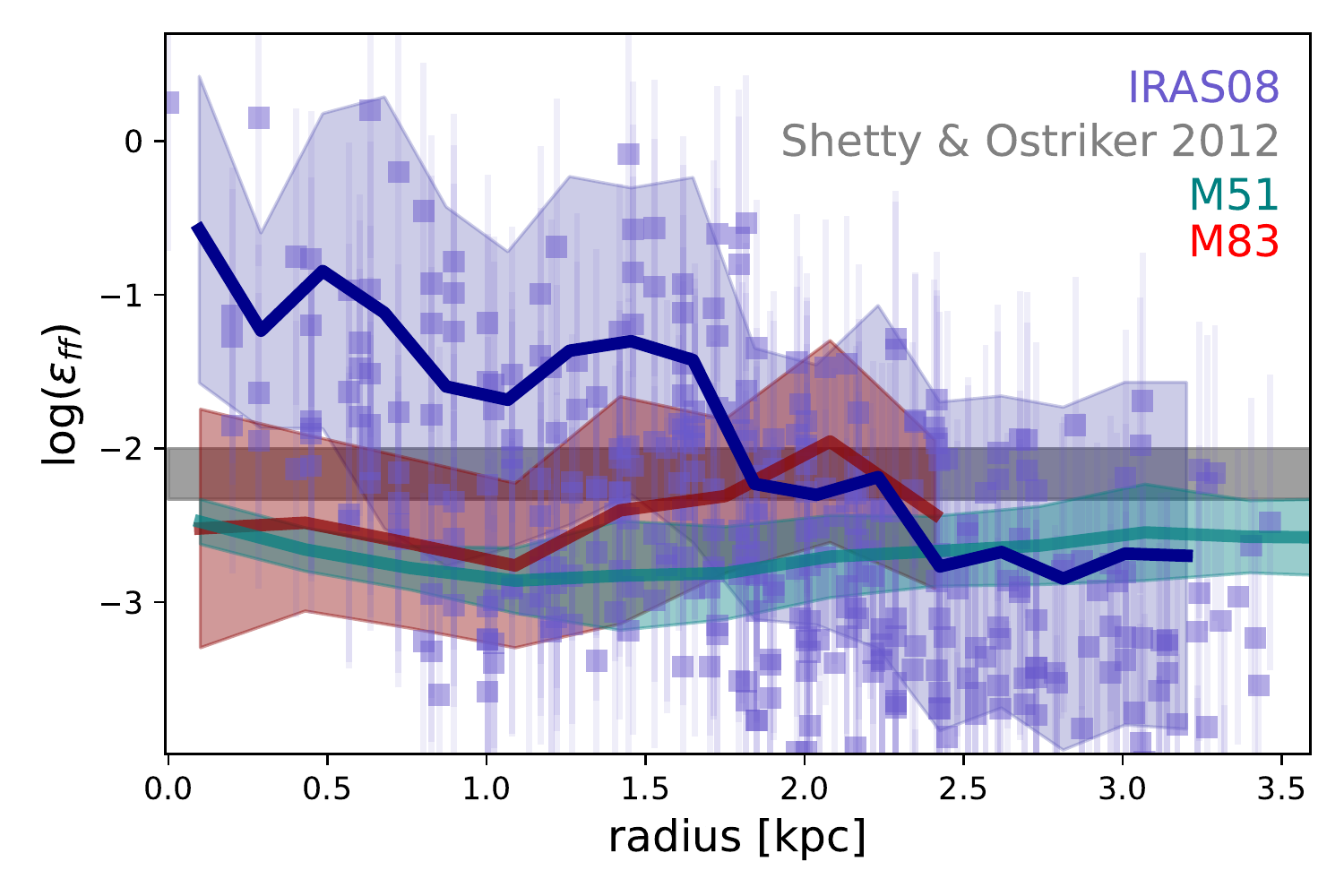}
\end{center}
\caption{ The radial distribution of star formation efficiency per free-fall time, $\epsilon_{ff}$, in IRAS08 for the circular cloud model. As before, the solid line represents the running average in 0.5~arcsec bins, and the shaded region represents the RMS logarithmic scatter. As with $t_{dep}$ there is a strong gradient in $\epsilon_{ff}$ within this galaxy, in this case spanning multiple orders of magnitude. As discussed above, this gradient cannot be accounted for by variation in the disk thickness. \label{fig:leffprof} } 
\end{figure}

Those methods that estimate the scale-height with $\sigma$ and $\Sigma_{gas}$ both show an increase in $t_{ff}$ by a factor of 3$\times$ from the center of IRAS08 to the outskirts, whereas the circular cloud model finds an increase of roughly $\sim1.7\times$. To study $\epsilon_{ff}$ we adopt $h_z$ as described in \cite{Wilson2019} as this formulation is specifically derived for star-bursting environments, like IRAS08. 

This prescription for disk thickness is 
\begin{equation}
    h_z \approx 0.2 \frac{\sigma^2}{\pi G \Sigma_{mol}}
\end{equation}
(adopted from Equation 7 in \citealp{Wilson2019}). The factor of 0.2 in front takes into account sources of pressure support from magnetic and cosmic ray sources, as described in \cite{kim2015}. It has been scaled from the value in \citealp{Wilson2019} for the fraction of gas-to-total mass in IRAS08.  It also considers non-local sources of gravity, as in the vertical component of the three-dimensional gravitational acceleration toward the inner part of the galaxy. It is here that we assume the higher surface density environment, which is appropriate for IRAS08. For a full description of their derivation see \cite{Wilson2019}.

Since the free-fall time varies as $h_z^{1/2}$, even in the most extreme limit, this particular assumption could not account for more than a 20\% change in the free-fall time. In Fig.~\ref{fig:b200} we show the radial profile of $\Sigma_{mol}/\sigma^2$, which scales inversely with the scale-height. This quantity is discussed in detail in \cite{Leroy2017} in a study of nearby spiral galaxy, M~51. As we show in the figure, IRAS08 has significantly lower values of $\Sigma_{mol}/\sigma^2$ than what is found in M~51, typically at the order of magnitude level. This implies that under similar assumptions about the calculation of $h_z$, IRAS08 has a thicker disk.

 We can use the result in Fig.~\ref{fig:b200} to estimate the variation in the disk thickness of IRAS08. There is considerable point-to-point scatter, but the radial averages are fairly flat. The average value of $\Sigma_{mol}/\sigma^2$ decreases by $\sim$0.4~dex from the highest value at $\sim$0.6~kpc to the region around $2\times R_{1/2}$ (2~kpc). 

We remind the reader that the variation the free-fall timescale is related to the square root of $h_z$, $t_{ff}\propto h_z^{1/2}$, which means that the changes in scale height of order $2\times$ only impact $t_{ff}$ and $\epsilon_{ff}$ by $\sim$0.15~dex. Combining this with 
alternatives to the adopted prescription, we 
estimate that choices for deriving $h_z$ in IRAS08 generates a systematic uncertainty of order $\pm$0.2~dex on $\epsilon_{ff}$.

In Fig.~\ref{fig:tff} we show that $t_{ff}$ increases from $\sim$3-5~Myr in the galaxy center to values of order 10-15~Myr at the 90\% radius, roughly a factor of $3\times$ increase. Note the low values of $t_{ff}$ at $\sim$1~kpc and 2~kpc are associated to peaks, likely clumps of CO gas.



\subsection{Distribution of $\epsilon_{\MakeLowercase{ff}}$ in IRAS08} 
In Fig.~\ref{fig:leffprof} we compare the radial distribution of $\epsilon_{ff}$ to local spiral galaxies (M~83 \citealp{Hirota2018}   \& M~51 \citealp{Leroy2017}). We find in IRAS08 that there is a very strong correlation of galactic radius with $\epsilon_{ff}$. At large radii in IRAS08 $\epsilon_{ff}$ is similar to what is found in local spirals, $\epsilon_{ff}\approx 0.3\%$. Inside $R\sim2$~kpc the profile of $\epsilon_{ff}$ shows a strong decrease of $\epsilon_{ff}$ with radius, that is not matched in either local spiral. The full range of variation of $\epsilon_{ff}$ in M~51 and M~83 is at the $\pm$0.3~dex level, where IRAS08 experiences a difference of roughly 2~orders-of-magnitude from the center to the outer disk.

We note the caveat that our study and that of \cite{Hirota2018} (M~83) use H$\alpha$ to trace SFR, while \cite{Leroy2017} (M~51) uses total IR luminosity. This may introduce a bias in derived $\epsilon_{ff}$ values. Different tracers reflect different time-scales of star formation, though this is very unlikely to account for the multiple order of magnitude difference that we observe \citep[for review][]{kennicuttevans2012arxiv}.  CO-to-H$_2$ conversion is another important systematic uncertainty. Using a starburst $\alpha_{CO}$ would have the effect of making a significant fraction of lines-of-sight in the central $\sim$50\% of the galaxy reach $\epsilon_{ff}\approx100\%$.  Alternatively, $\alpha_{CO}$ could vary with local surface density, as described in the results on $t_{dep}$. In IRAS08 this would yield act to steepen the correlation in both $t_{dep}$ and $\epsilon_{ff}$, with disk-like efficiency at large radius and extreme efficiencies of order $\epsilon_{ff}\approx100\%$ in the central few kpc. Our constant Milky Way $\alpha_{CO}$ assumption is therefore a conservative estimate of the value and gradient of $\epsilon_{ff}$ in IRAS08. 

Very high $\epsilon_{ff}$ are rare in the local Universe, and are more similar to those observed in super-star clusters (SSCs) in local galaxies \citep{Turner2015}. There are only a handful of observations capable of measuring the efficiency in SSCs. The center of NGC~253 offers a rich starbust environment in which \cite{Leroy2018} observes an overall efficiency of star formation of $\sim$50\% in SSCs and an efficiency per free-fall time that is similar to the central kiloparsec of IRAS08 (when using a Milky Way $\alpha_{CO}$), $\epsilon_{ff}\sim10$\%. Similar results are found in the SSC Mrk~71A \citep{Oey2017}.  Indeed, \cite{otifloranes2014} carries out a multiwavelength study of IRAS08, and finds that the center is consistent with containing SSC.

Recently, \cite{Utomo2018} measured $\epsilon_{ff}$ at a similar resolution to ours in a sample of local spiral galaxies from PHANGs. They do not study the radial variation of $\epsilon_{ff}$, but do provide mean and RMS values. The key difference between  $\epsilon_{ff}$ measured in IRAS08 to that of the local spirals in \cite{Utomo2018} is that IRAS08 has a much larger RMS, and that that spread is skewed to higher $\epsilon_{ff}$. The RMS of $\epsilon_{ff}$ is 0.76~dex in IRAS08 compared to 0.25~dex in the disks from PHANGS. 

Using a standard, unweighted, median yields a $<\epsilon_{ff}> \approx 0.3\%$, similar to \cite{Utomo2018}.  An unweighted average implicitly weights the average $\epsilon_{ff}$ to the larger radius regions, where there are more lines-of-sight. An average that is weighted by SFR yields $<\epsilon_{ff}>_{SFR}\approx8-10\%$ and weighting by CO flux yields $<\epsilon_{ff}>_{CO}\approx1\%$.

The weighted averages imply that, like $t_{dep}$, higher $\epsilon_{ff}$ in IRAS08 is more strongly correlated with $\Sigma_{SFR}$ than with $\Sigma_{mol}$. We find that the Spearman rank correlation coefficient of $\epsilon_{ff}$ with $\Sigma_{H\alpha}$ is stronger ($r\approx 0.60$) than with $\Sigma_{mol}$ ($r\approx0.45$). We note that the strong correlation of $\epsilon_{ff}$ with $\Sigma_{SFR}$ is also different from what is observed in M~51 and M~83. There is not a statistically significant correlation of $\epsilon_{ff}$ with $\sigma$ ($r\approx0.15)$. 




\begin{figure*}
\begin{center}
\includegraphics[width=18cm]{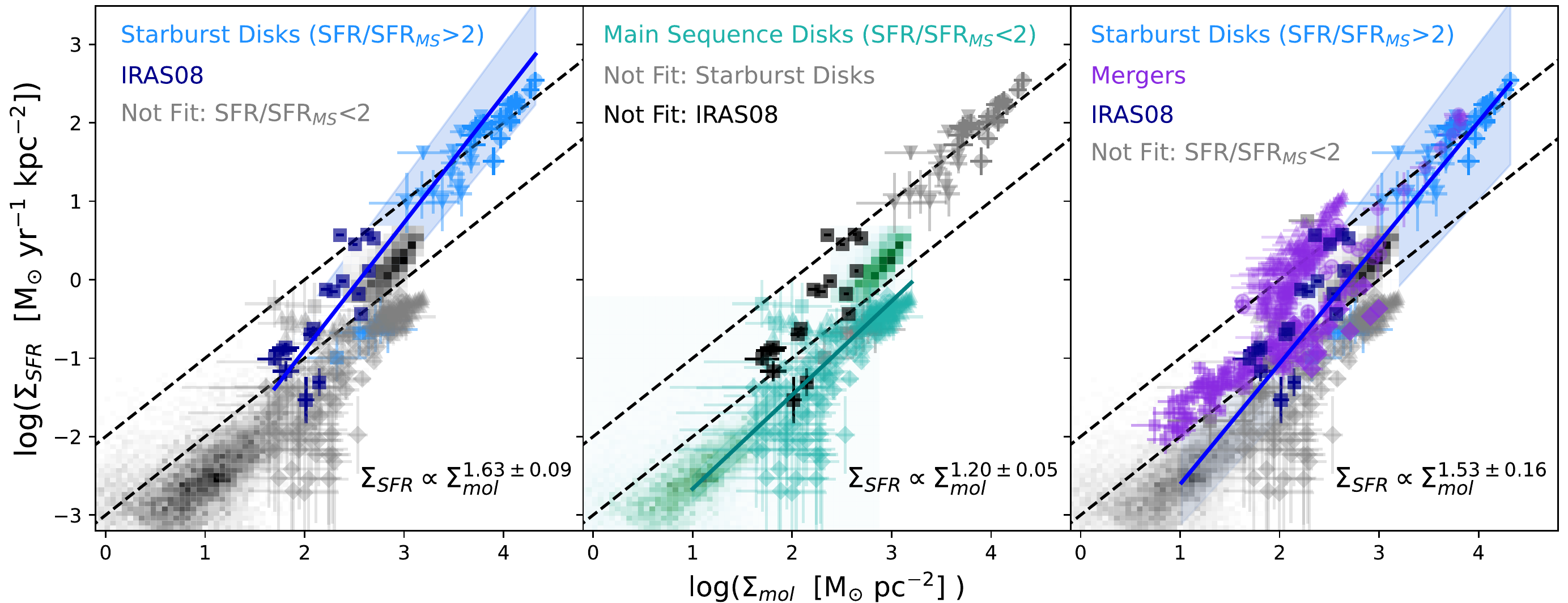}
\end{center}
\caption{The resolved (kiloparsec-scale) relationship between $\Sigma_{SFR}$ and $\Sigma_{mol}$ is shown above. In each panel we fit a subset of targets, which are indicated as having blue (left), green (middle) or blue \& violet (right) colors. In all panels the points that are grey color are not included in the respective fit. The powerlaw of each fit is given in the lower right corner of each panel. The {\bf left panel} shows a fit to the blue points (light and dark), which are rotating disks that have SFR that is at least 2$\times$ higher than the main-sequence value for their respective mass, including IRAS08. The {\bf middle panel}  shows a fit to  the green points, which are disks -- at both low and high redshift-- that are on the main-sequence.
The {\bf right panel} shows a fit to both the blue and violet points, which are star bursting disks (blue) and those systems that from morpho-kinematic analysis are likely to be significantly affected by their merging (violet).  The $\sim1\sigma$ scatter around the  weighted average in all panels is shown as a shaded region. The full range in $\Sigma_{SFR}-\Sigma_{mol}$ parameter space has significant complexity. There are two sequences, that appear to be more closely connected to distance to the main-sequence than morpho-kinematic state (i.e. merging or rotating). We point out the very considerable scatter of all points in the range $\Sigma_{mol}\approx 10^2-10^3$~M$_{\odot}$~pc$^{-2}$ implies a lack of single $\Sigma_{mol}$ at which the separation of starburst occurs. 
\label{fig:sflaw}}
\end{figure*}


\section{The $\Sigma_{SFR}-\Sigma_{mol}$ scaling relationship at kiloparsec resolution}

To place IRAS08 into context with other $z>1$ and high $\Sigma_{mol}$ star forming galaxies we use the relationship between SFR surface density and gas mass surface density at roughly kiloparsec scale resolution. We highlight 2 results from IRAS08. First, it further illustrates that there is not a simple cut-off at $\Sigma_{mol}\sim 100-200$~M$_{\odot}$~pc$^{-2}$ separating all starburst from non-starburst galaxies in this parameter space; the separation is more complex. Secondly, despites its highly variable $\epsilon_{ff}$, IRAS08 is not an outlier, implying that variable $\epsilon_{ff}$ could be common in $z>1$ starburst galaxies.

\vskip 5pt
This relationship is typically characterised by a power-law where
 \begin{equation}
\Sigma_{SFR} = A_{\ } \Sigma_{mol}^N,  \label{eq:sflaw}
\end{equation}
where $A$ and $N$ are fitted parameters \citep{kennicutt98,kennicuttevans2012arxiv}. The power-law slope, $N$ in Equation~\ref{eq:sflaw}, has been interpreted as a constraint on physical models of star formation \citep[e.g.][]{ostriker2010,krumholz2017,Elmegreen2018,Semenov2019}, and is therefore of particular interest. There is a large amount of literature on this correlation across a range of surface brightness, for recent reviews see \cite{Tacconi2020arxiv} and also \cite{hodge2020arxiv}.

To compare IRAS08 to measurements of $t_{dep}$ for galaxies in the literature we must degrade the resolution of our data cube to a similar resolution (800~pc) and remeasure the integrated intensity map. We note the well-known biases of of how spatial scale affects $\Sigma_{mol}-\Sigma_{SFR}$ relationship  are described in literature \citep[e.g.][]{calzetti2012,leroy2013,Kruijssen2014}. The comparison sample includes the following galaxies: local disks from the HERACLES project \citep{leroy2013}; local Universe wide-seperation interacting galaxies NGC~232 and NGC~3110 \citep{espada2010}; local Universe advanced merging galaxies VV~114 \citep{Saito2015}, NGC~1614 \citep{Saito2016} and the Antenna System \citep{Bemis2019}; rotating $z=1-4$ galaxies EGS~13011166 \citep[z$\approx$1.5;][]{genzel2013}, GN20 \citep[$z\approx4$;][]{hodge2015}, AzTEC-1 \citep[$z\approx4$;][]{tadaki2018}, SHIZELS-19 \citep[$z\approx1.5$;][]{Molina2019}, SDSS~J0901+1814 \citep[$z\approx 2.3$;][]{Sharon2019} and $z>1$ systems that are more consistent with being merging galaxies HATLAS J084933 \citep[$z\approx2.4$;][]{gomez2018arxiv}, ALESS67.1 \citep[$z\approx2.1$;][]{chen2017aless}. Due to the biases in sampling size discussed above, we refrain from plotting measurements of entire galaxies, and also refrain from plotting measurements less that 0.4~kpc. 


To measure the powerlaw relationship between star formation and molecular gas mass surface density we carry out Monte Carlo fit of the data sets to Equation~\ref{eq:sflaw} using Ordinary Distance Regression in the {\em Python} package {\em scipy}. We weight the data so that each galaxy has equal impact on the fit. Data points are also weighted by the measurement uncertainty. For each fit, we run 1000 realizations in which the data points are shifted randomly within 0.3$~dex$ in both SFR and gas mass to account for systematic uncertainties. Increasing the size of the range in which we shift points, within reasonable limits, only has a minor impact on the derived powerlaw. The reported powerlaw slope and scale factor are the median values from the iterations, and the uncertainty is the 1$\sigma$ scatter around this value.  We test our method first by fitting the HERACLES data only in the range 10-100~M$_{\odot}$~pc$^{-2}$, and we recover a powerlaw of $N=0.97$, which is very similar to values measured in \cite{leroy2013}. Results of fits are shown in Fig.~\ref{fig:sflaw} and tabulated in Table~\ref{table:fits}.

For the fits in Fig.~\ref{fig:sflaw}, we group galaxies together based on coarse galaxy properties including: morpho-kinematic state (i.e. disks versus mergers) and  distance to the star forming main-sequence. We use the redshift evolution of the main-sequence as defined in \cite{whitaker2012}. We define the distance from the main-sequence as the ratio of the observed star-formation rate to the main-sequence star formation rates ($\delta MS\equiv SFR/SFR_{MS}$). When multiple estimates of star formation rate were available we opt for those made from ionized gas for consistency.

We find  $N\approx1.2$ for main-sequence galaxies. This is marginally steeper than what is found for fits to HERACLES disks alone ($N\approx0.97$), but similar to the steep slope found by \cite{genzel2013} for the $z\sim1.5$ main-sequence galaxy. Galaxies identified as having starbursts show powerlaws with $N\sim 1.6$. We find that separating star-bursting galaxies between those suspected of being mergers and those that are not mergers has very little impact on the power-law slope derived from fitting Equation~\ref{fig:sflaw}. IRAS08 is in general agreement with the star-bursting sequence. It has a range $\Sigma_{mol}\sim 100-400$~M$_{\odot}$~pc$^{-2}$ that overlaps with both the starbursting and non-starbursting sequence. On its own it has a steeper power-law ($N\sim1.8-2.0$). We note that because IRAS08 is compact (R$_{1/2}\sim 1$~kpc), when resampled to lower resolution there are only a few independent data points, and thus the fit to only IRAS08 has significant uncertainty. Moreover, we reiterate from Section~4 that the lower resolution averages soften the gradient in $t_{dep}$. We do not find it useful to study higher spatial resolution relationship of $\Sigma_{SFR}-\Sigma_{mol}$ for two reasons. First, our purpose of studying the $\Sigma_{SFR}-\Sigma_{mol}$ relationship is for comparison to other galaxies, and it is well established that such comparisons must be carried out on a comparable spatial scale \citep{leroy2013}. Secondly, it is clear from Fig.~\ref{fig:tdep_prof} that there is considerable scatter of individual beams. Fits using standard methods, as we use here, are heavily dominated by uncertainty.

We find that there is not a simple threshold in behavior at a single $\Sigma_{mol}$, as suggested previously in the literature \citep{bigiel2008}. IRAS08 has a similar range of $\Sigma_{mol}$ and both EGS~13011166 and SHIZELS-19 but has a significantly steeper slope than both. The key parameter that distinguishes galaxies on the two tracks, is distance from the main-sequence. Similar arguments are discussed in \cite{Tacconi2020arxiv} describing global depletion times. We argue $\Sigma_{mol}$ on its own should not be used to discriminate between the two sequences in the $\Sigma_{SFR}-\Sigma_{mol}$ relationship. 

\begin{deluxetable}{lcc}
\tablewidth{0pt} \tablecaption{Fits to $\Sigma_{SFR} =A\Sigma_{mol}^N$}
\tablehead{ \colhead{Category} & \colhead{log($A$) } & \colhead{$N$} \\ 
}
\startdata
$\delta$MS$>2$ disks only & -4.14$\pm$0.27 & 1.63$\pm$0.09\\
$\delta$MS$>2$ All & -4.15$\pm$0.38 & 1.53$\pm$0.16\\
$\delta$MS$<2$ disks only & -3.85$\pm$0.12 & 1.20$\pm$0.04\\ \\
\hline \\
\enddata
\label{table:fits} 
\end{deluxetable}

In spite of its large high variability in $\epsilon_{ff}$ in IRAS08, Fig.~\ref{fig:leffprof}, the galaxy is not a significant outlier from other starbursting systems. It is in fact less extreme in $\Sigma_{mol}$ and $\Sigma_{SFR}$, by an order-of-magnitude, than AzTEC-1 \citep{tadaki2018} and GN20 \citep{hodge2015}, and has similar gas and SFR surface densities as $z\approx 1-2$ galaxies, e.g  SDSS~J0901+1814 \citep{Sharon2019}. This implies that such extreme values of $\epsilon_{ff}$ could be wide-spread in starbursting galaxies of the distant Universe. Given its similarity to other star-bursting galaxies in the subsequent section we consider the implications of the results in Figs.~\ref{fig:leffprof} and \ref{fig:sflaw} on star formation models in the literature.

\section{Discussion}
\subsection{Implications for Galaxy-Scale Star Formation Theories}
The combined results in Fig.~\ref{fig:leffprof} and Fig.~\ref{fig:sflaw} allow for a direct comparison to a number of theories for how star formation evolves in galaxies.  We note that the extreme nature of star formation in IRAS08 does not preclude comparison to these theories, as almost all explicitely discuss the relevancy for star-burst regime  \citep[e.g.][]{shetty2012,krumholz2017,faucher2013,Elmegreen2018}.

There is a clear tension between our results in  Fig.~\ref{fig:leffprof} and those theories that assume or derive a constant star formation efficiency per free fall-time \citep[e.g.][]{krumholz2012,salim2015,Elmegreen2018}. In such theories $\Sigma_{SFR}=\epsilon_{ff}/t_{ff}\Sigma_{gas}$, and that $\epsilon_{ff}$ is constant, but $t_{ff}$ varies. In IRAS08 we find that the opposite is true. We find very little variation across the disk in $t_{ff}$ (of the order of a factor of a few), but a variation in $\epsilon_{ff}$ of a factor of $\sim$50$\times$. Moreover, recent observations \citep{Fisher2019} have established that on galaxy scales there is an inverse relationship of $t_{dep}\propto \sigma_{gas}^{-1}$, which is opposite the prediction of constant star formation efficiency models.

\cite{Elmegreen2018} reviews how different physical regimes may lead to different powerlaws in the Kennicutt-Schmidt diagram. They show that for a disk 
\begin{equation}
\Sigma_{SFR} \propto \epsilon_{ff} h_z^{-1/2}\Sigma_{gas}^{3/2}. 
\label{eq:consthz}
\end{equation}
While we do see that in starburst galaxies $\Sigma_{SFR}\propto\Sigma_{mol}^{3/2}$, in order for Equation~\ref{eq:consthz} to hold for IRAS08 the change in $\epsilon_{ff}$ would need to be canceled by a greater change in $h_z$. Under the assumption that  variation in the scale height of a disk is traced by $\sigma^2/\Sigma_{mol}$, we would need a factor of $\sim$200$\times$ to cancel the change in $\epsilon_{ff}$. We find however, in Fig.~\ref{fig:b200} that the ratio $\Sigma_{mol}/\sigma^2$ only changes by a factor of $\sim$2$\times$ across the disk.  This implies that 
$h_z$ is relatively constant, while $\epsilon_{ff}$ systematically increases with $\Sigma_{SFR}$ by orders-of-magnitude. This formulation, therefore, appears inconsistent with observations of star formation in IRAS08.

A third class of models predict that star formation is regulated by feedback from newly formed stars \citep[e.g.][]{ostriker2010,shetty2012,kim2013,faucher2013}. \cite{shetty2012} specifically investigates the regime of maximally starbursting disks, and is therefore applicable to our observations, they argue that 
\begin{equation}
\Sigma_{SFR}\propto (p_*/m_*)^{-1}\Sigma_{gas}^{2}
\label{eq:feedback}
\end{equation}
The quantity $p_*/m_*$, is the momentum input into the ISM from supernova per mass of new stars formed, sometimes referred to as the ``feedback efficiency" \citep{kim2013}. For comparison to IRAS08 this prediction has the advantage of not simultaneously depending on both $\epsilon_{ff}$ and $t_{dep}$.  As we show in Fig.~\ref{fig:leffprof}, the simulations of \cite{shetty2012}, which incorporate these concepts, find that, even at high $\Sigma_{mol}$, the maximum $\epsilon_{ff}$ is $\sim$1\%, and thus a factor of $\sim$50$\times$ too low to describe the variability in IRAS08.  

In order for Equation~\ref{eq:feedback} to match the observations in Fig.~\ref{fig:sflaw} $p_*/m_*$ would have to increase with $\Sigma_{gas}$ (or $\Sigma_{SFR}$), at roughly $p_*/m_* \propto \Sigma_{gas}^{1/2}$, but for starburst galaxies only. The feedback efficiency is typically derived, or adopted, as a constant in star formation theories \citep[e.g.][]{ostriker2010,faucher2013,krumholz2017}.

\cite{Fisher2019} shows that constant feedback efficiency models have trouble describing the global relationships for both $\Sigma_{SFR}$ and gravitational pressure \citep[also see][]{Sun2020,Girard2021}. If the value of $p_*/m_*$ were to increase with either $\Sigma_{SFR}$ or $\Sigma_{mol}$, as described above, this would alleviate the discrepancies at high $\Sigma_{SFR}$ with both the correlations of $\sigma-t_{dep}$ and $\Sigma_{SFR}$ versus hydrostatic pressure. 

Though debate still exist \citep{kim2017}, some simulation work finds a significant increase in $p_*/m_*$ is possible in regions of higher star formation rate surface density, due to the effect of clustered supernova driving more efficient feedback \citep{gentry2017,gentry2018arXiv,Martizzi2020}. Moreover, simulations of outflows in starbursting systems like M82 or $z\sim2$ galaxies find that constant feedback efficiency models are not capable of reproducing the high velocity winds, whereas clustered supernova are \citep{Fielding2018}. We note that similarly in IRAS08 \cite{Chisholm2015} observes very high velocity winds, $v_{out}\sim 1000$~km~s$^{-1}$, using UV absorption lines. IRAS08 in fact has among the most rapid outflows in their sample of 48 local Universe galaxies. The $\Sigma_{SFR}$ clumps we observe in IRAS08 seem like an ideal location for the effects of clustered supernova. Such a change could also act to increase the {\em observed} $\epsilon_{ff}$ as more efficient feedback would decrease $\Sigma_{mol}$. 

Alternative to varying the feedback efficiency others argue that star formation is regulated by a combination of feedback and dynamical disk stability \citep{faucher2013,krumholz2017}.  In these theories the $\Sigma_{SFR}-\Sigma_{gas}$ relationship depends on both $p_*/m_*$ and Toomre $Q$, such that 
\begin{equation}
\Sigma_{SFR}\propto Q (p_*/m_*)^{-1}\Sigma_{gas}^{2}.
\end{equation}
To first approximation, this is consistent with a picture of galaxy evolution in which main-sequence galaxies have high values of $Q$ and starburst galaxies, which may be experiencing a violent disk instability, have low values of $Q$. This would then explain why there are multiple sequences in Fig.~\ref{fig:sflaw} at large $\Sigma_{mol}$, and is consistent with what we observe in Fig.~\ref{fig:qprof} for IRAS08, as well as AzTEC-1 \citep{tadaki2018}. We note that testing the $Q$ dependance on high-$z$ galaxies is more difficult than it seems, as  systematic uncertainties can have very large effects on both how the velocity dispersion is measured and the molecular gas is estimated \citep{Girard2021}. We note that \cite{Girard2021} shows that when SFR is compared to molecular gas velocity disperions, instead of ionized gas, the mixed feedback-transport model from \cite{krumholz2017} does not agree with data. 

\cite{faucher2013} creates a similar feedback-regulated model of star-formation, which incorporates dynamical regulation of the disk, such that $Q\approx 1$ and in their model $\epsilon_{ff}$ is free to vary.  They find a range of $\epsilon_{ff}$ at all $\Sigma_{mol}>100$~M$_{\odot}$~pc$^{-2}$. They also make predictions for global properties like $\Sigma_{gas}$, $f_{gas}$ and $\sigma$. The predict a disk averaged $\epsilon_{ff}$ as high as 30\% for a galaxy with properties like IRAS08. Though it is not clear from their model if there is a systematic variation of $\epsilon_{ff}$ like we see in IRAS08. 

Similar to these gravitationally based prescriptions, there is a long known result \citep{kennicutt98araa} that for total gas mass in galaxies normalising the gas mass by the orbital time-scale creates a linear correlation that galaxies obey well, such that $t_{dep}\propto t_{orb}$, where the orbital timescale is defined as $t_{orb}=2\pi R/V$. This amounts to stating that galaxies convert a constant fraction of gas into stars per orbit. In IRAS08 both $t_{dep}$ and $t_{orb}$ become larger with radius, causing a positive correlation.
The change in $t_{orb}$, however, is insufficient to account for the two orders-of-magnitude change in $t_{dep}$.  We find that in the central kiloparsec in IRAS08 $t_{orb}\approx30-50$~Myr, rising to $\sim150$~Myr in the outer disk, an increase of a factor of 5.

We summarise how our results compare to star formation: \\
$\bullet$ \underline{Constant Star Formation Efficiency Models:} 
Our observations of IRAS08 are inconsistent with theories in which the star formation efficiency is held constant \citep{krumholz2012,salim2015}. Constant star formation efficiency models also fail to recover the observed relationship of $t_{dep}\propto \sigma^{-1}$ \citep{Fisher2019}. 

\noindent $\bullet$ \underline{Feedback Regulated Models:} Models in which star formation is regulated only by the balance of feedback with local gravity \citep[e.g.]{shetty2012}, excluding large scale galactic flows, could explain the properties of galaxies like IRAS08 only if more freedom is given to both star formation efficiency, and especially if the efficiency of feedback is allowed to be higher in higher SFR surface density regions, perhaps due to supernova clustering. 

\noindent $\bullet$ \underline{Mixed Feedback+Toomre Regulation:} Models in which feedback effects are mixed with disk self-regulation via Toomre instabilities appear most consistent with our observations \citep{faucher2013,krumholz2017}. Such models have a built-in explanation for the multiple sequences in the $\Sigma_{SFR}-\Sigma_{mol}$ relationship. Moreover, the model of \cite{faucher2013} does allow for larger disk averaged $\epsilon_{ff}$. Though these models do not, as yet, give testable predictions for the systematic variation in $t_{dep}$ and $\epsilon_{ff}$ within IRAS08, and the correlation of SFR-$\sigma_{mol}$ for molecular gas velocity dispersions does not match the data in samples of both low- and high-z galaxies \citep{Girard2021}. 

It is important to emphasize that our findings are based on only one galaxy. The stark differences from what we observe and commonly accepted theories of star formation seem to strongly argue for more observations of resolved maps of molecular gas in high $\Sigma_{SFR}$ disk galaxies. Whether that be directly at $z>1$ or with analogue samples such as DYNAMO \citep{fisher2017mnras}, it is now needed to determine if strong gradients in $\epsilon_{ff}$ are common in this mode of star formation.



\subsection{Possible dynamical drivers of $\MakeLowercase{t_{dep}}$ gradient}


If we assume that the gas flow which was responsible for the variation in $\epsilon_{ff}$ and $t_{dep}$ is quasi-stable on a similar timescale as $t_{dep}$ (100-500~Myr) we can expect that evidence for such an inflow may still present in the galaxy. It is particularly interesting to consider the similarity between IRAS08 and blue-compact galaxies at $z\approx 1-2$, which are thought to experience very rapid inflows as an important component of galaxy evolution \citep{dekel2014,tacchella2016}. 

\vskip 5pt
We first look for independent evidence of inflow using the metallicity profile. Flat metallicity gradients are frequently interpreted as indicators of gas inflows within galaxies \citep{Kewley2010,Jones2013}. The reason is straightforward. The higher density of star formation in the galaxy center (Fig.~\ref{fig:tdep_prof}) should pollute the ISM faster, and therefore in order to maintain a flat metallicity profile the center must be replenished with less metal rich gas. 

\begin{figure}
\begin{center}
\includegraphics[width=0.5\textwidth]{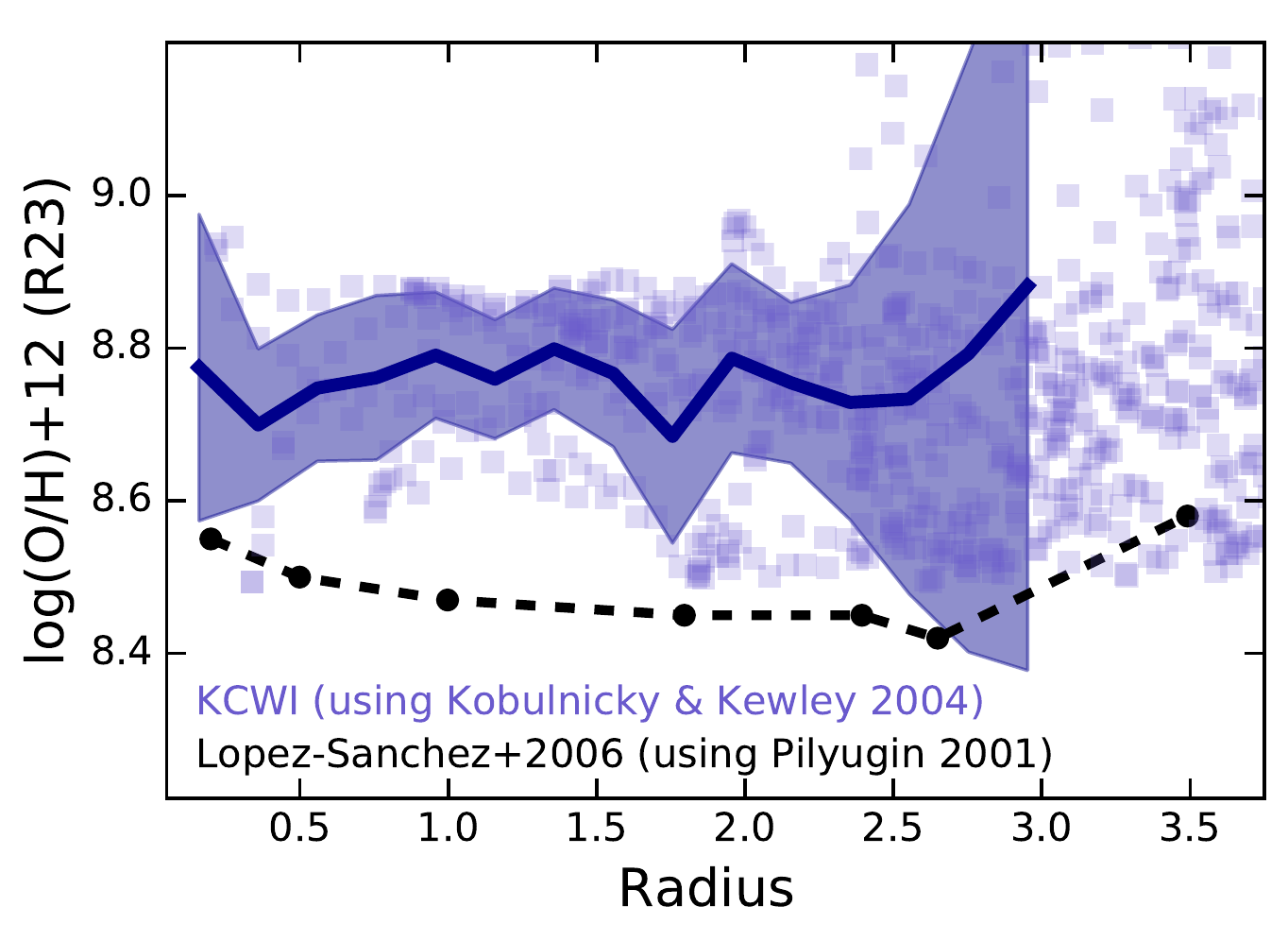}
\end{center}
\caption{The metallicity profile determined via the R23 method as described in \cite{kobulnicky2004}. Note that we do not plot error bars because the typical strong-line measurements across this region of the galaxy have $S/N\sim100$, as the data was intended for analysis of the fainter features in outflows. A future paper Reichardt-Chu et al. {\em submitted} provides an in depth analysis of the KCWI dataset. We also show the metallicity profile measured by \cite{Lopez2006} as a black dashed line.  Across the disk of IRAS08 we observed an essentially flat metallicity profile. Shallow, or even negative, metallicity gradients are widely interpreted as indicating gaseous inflows across the disk. \label{fig:metal}} 

\begin{center}
\includegraphics[width=0.5\textwidth]{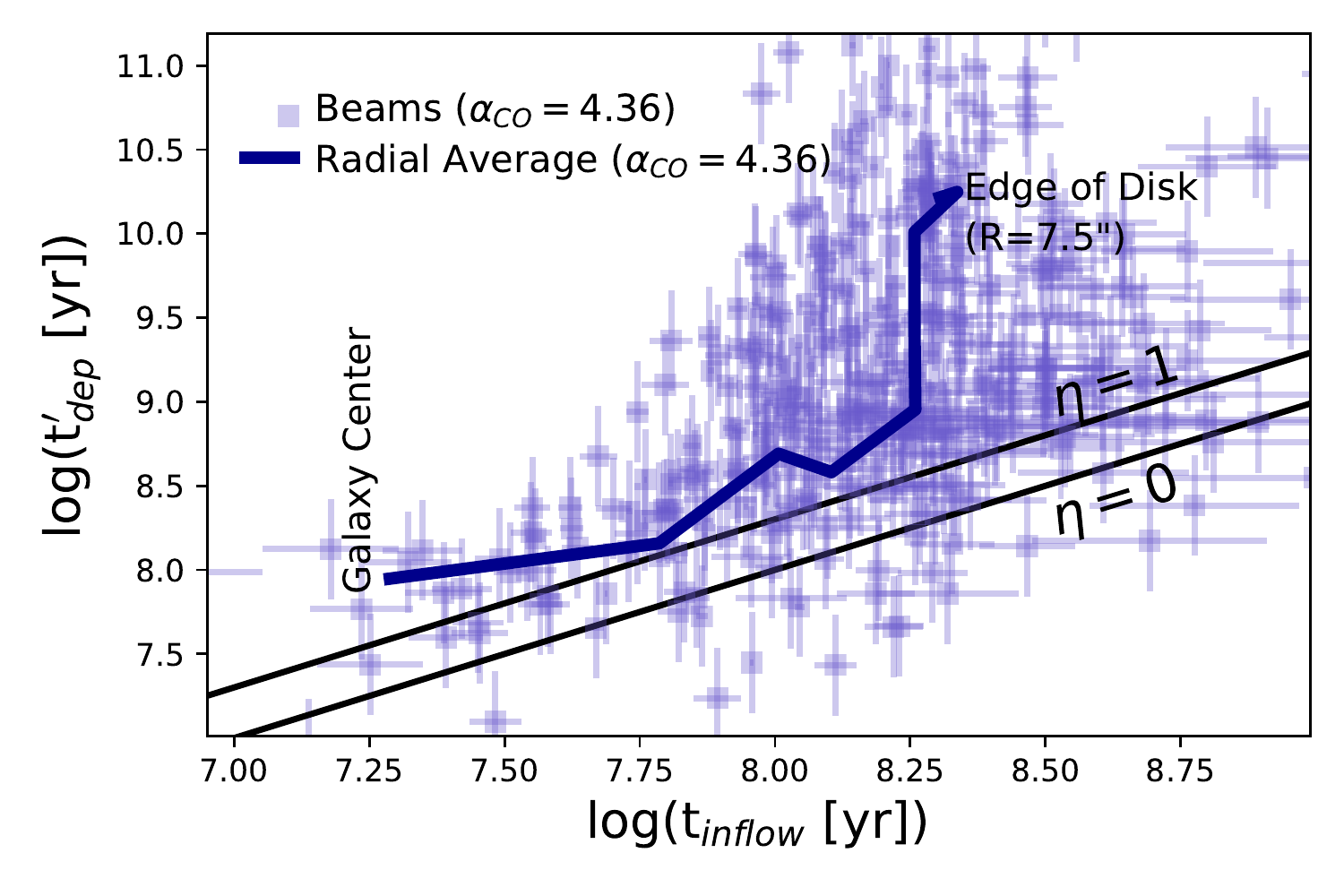}
\end{center}
\caption{The molecular gas depletion time, $t_{dep}\equiv \Sigma_{mol}/\Sigma_{SFR}$ is plotted against the inflow timescale, $t_{inf}$ (Equation~\ref{fig:tinf}), as determined from \cite{dekel2014}. The squares represent individual beams in the CO(2-1) map. The error bar indicates the uncertainty from $\alpha_{CO}$. The thick, blue solid line represents the average determined as a function of radius in increments of 1~arcsec ($\sim0.8$~kpc). The black line represents the line of equality assuming a mass-loading factor of $\eta=0$ (bottom) and $\eta=1$ (top). In IRAS08 $t_{dep}$ varies by 2~orders-of-magnitude from the outer disk to the inner kiloparsec, however the average does not drop below the inflow timescale.     
\label{fig:tinf}}

\end{figure}

The metallicity profile, shown in Fig.~\ref{fig:metal}, is consistent with an inflow of gas toward the galaxy center.  The blue points indicate our measurements using the R23 method from \cite{kobulnicky2004} with KCWI data, and the black points show the measurement from \cite{Lopez2006}, which uses the \cite{pilyugin2005} calibration for the same emisison lines. There are very well known offsets between the metallicity of different calibrations \citep[for review see][]{Kewley2019Review}. The gradient of our measurement and \cite{Lopez2006} is similarly flat across the disk. A typical, massive spiral galaxy has a metallicity gradient, using the R23 method, of order -0.4~dex~$R_{25}^{-1}$ in log(O/H), see \cite{Ho2015} and reviewed in \cite{Bresolin2017}. For an exponential disk $R_{25}$ is roughly equivalent to the 90\% radius. We measure a 90\% radius of star light, using HST F550M image, of 3~kpc. Across this range the metallicity profile of IRAS08 shows no decrease at all, consistent with inflow scenarios.   

We consider 3 possible mechanisms for gas inflows: galaxy wide ``violent" disk instability, a distant interaction with a neighboring galaxy, and the bar in the galaxy center. 

\subsubsection{Violent Disk Instability as inflow driver}

Typically, violent disk instabilities as drivers of inflow are connected to the phenomena of ``wet compaction" in $z\approx1-2$ galaxies. In this scenario rapid inflows of gas quickly build bulges in starbursting disks \citep[see discussion in][]{Zolotov2015}. Observations and simulations of high-$z$ galaxies associate the phenomenon of compactions with blue compact galaxies. As described above, IRAS08 has historically been treated as a rare local analogue of luminous blue-compact galaxies in the distant Universe \cite{Lopez2006,ostlin2009}. We have also measured a low Toomre~Q in IRAS08. We therefore consider the possibility that a similar phenomena is dominating the inflow of gas in IRAS08.

\cite{Zolotov2015}, and also recently \cite{Dekel2020arXiv}, describe the properties of simulated galaxies experiencing these phenomena. The critical properties are high specific SFR (SFR/M$_{star}$) and high stellar mass surface density in the central kiloparsec. In IRAS08 we observe a specific SFR in the central kiloparsec of $\sim$1.1~Gyr$^{-1}$ and a stellar mass surface density of $\Sigma_{*}=2.9\times10^{9}$~M$_{\odot}$~kpc$^{-2}$. These values place IRAS08 within the range of values for galaxies experiencing compaction in the simulations analysed by \cite{Zolotov2015}. In Fig.~\ref{fig:tdep_prof} we show that in IRAS08 the molecular gas surface brightness peaks at a radius of $\sim$0.5-0.7~kpc, and then declines at larger radius. \cite{Dekel2020arXiv} finds that such gas profiles are similar to pre-compaction or early-compaction galaxies. We note that there are no observations of blue-compact galaxies at $z>1$ with sub-kpc resolution, like our observations of IRAS08, it is therefore not possible to determine if such rings are common or not in blue compact galaxies thought to experience compaction.  

\cite{dekel2014} argue that wide-scale {\em violent disk instabilities} naturally drive inflows of gas, and make a testable predicitions for the internal distribution of $t_{dep}$. They develop a formalism in which the inflow timescale can be estimated from the assumption that (1) the kinematics are that of marginally stable/unstable disk, and (2) that energy gained from inflow is equal to the energy dissipated via turbulence. These assumption yield an inflow timescale, $t_{inflow}$ of 
\begin{equation}
t_{inflow}\approx 2_{} \frac{R}{V}_{} \left(\sqrt{2}\frac{\sigma}{V}\right)^{-2}.
\label{eq:tinf}
\end{equation}
If the galaxy kinematics are dominated by the violent disk instability then the inflow timescale is predicted to always be greater than the depletion time of the gas, $t_{dep}'$. Here we use a modified definition of gas depletion time, $t_{dep}'$ as 
\begin{equation}
t_{dep}'\equiv \frac{M_{gas}}{SFR_{\ }(\eta+1)},
\label{eq:tdep}
\end{equation} 
where $\eta$ is the outflow mass-loading factor. If $t_{dep}'\leq t_{inflow}$ then  gas will convert to stars before it reaches the galaxy center. The inflow timescale, as described in Equation~\ref{eq:tinf}, is only relevant to galaxies experiencing a violent disk instability.

The condition that $t_{dep}>t_{inflow}$, therefore, gives us a testable condition for consistency with the wet-compaction scenario for IRAS08. A system in which the gas inflow is driven by torques due to merging or accretion would not need to obey the condition $t_{dep}>t_{inflow}$ (where $t_{inflow}$ is derived from Equation~\ref{eq:tinf}) to maintain inward gas movement. 

We note that using Equation~\ref{eq:tinf} to estimate the inflow timescale makes an explicit assumption that the galaxy wide disk instability is driving the gas flow. In IRAS08 there is clearly a bar, and there is a significant amount of literature on the impact of bars, and associated resonances impacting gas inflows \cite[see reviews][]{kk04,athan05}. We will consider bars later as a driver of the gas inflow. 

In Fig.~\ref{fig:maps} the molecular gas is preferentially located in spiral arms, this may imply that the assumptions of linear Toomre instabilites are not applicable. Under this case the formulation of $t_{inflow}$ in Equation~\ref{eq:tinf} may not be correct. We can make approximate estimates of the impact of the nonlinearity, on the formulation of $t_{inflow}$. The presence of large clumps of gas with masses of order 10$^7$~M$_{\odot}$ implies that some instability must have recently existed. One possibility is that our measurements of $Q$ which focus on clumps under-estimate the $Q$ value of the disk gas. Simulations of clumpy galaxies in the non-linear regime find that $Q$ can be as high as $Q\sim1.8$ \cite{inoue2016}. Under very simplistic assumptions, this would have the effect of increasing $t_{inflow}$ by a factor of 3$\times$. Alternatively, the fact that that the molecular gas favours the spiral arms may imply that a the clumps are in response to a spiral arm instability, as described in \cite{inoue2018}. They show that many of the results of Toomre theory, have only minor corrections to clumpy spiral arm instabilties.


In Fig.~\ref{fig:tinf} we compare the molecular gas depletion time to the inflow timescale determined at each CO beam in IRAS08. The straight lines indicate the lines of equality for typical assumptions on the mass-loading factor of outflowing gas from star bursting regions \citep[e.g.][]{bolatto2013,Veilleux2020}. We show that while $t_{dep}$ decreases as a function of radius within IRAS08, very few of the measured beams have $t_{dep}'<t_{inflow}$. We also averaged both $t_{dep}$ and $t_{inflow}$ in radial bins of $\sim$0.8~kpc. In Fig.~\ref{fig:tinf} we show that the two timescales decrease in such a way that the $t_{inflow}$ is never less than the depletion time. This galaxy therefore satisfies the condition in \cite{dekel2014} for gas driven inflows by violent disk instabilities. 

The result in Fig.~\ref{fig:tinf} does not absolutely mean that the gas inflow in IRAS08 is driven by disk instabilities. Indeed as we have stated above the fact that the gas is preferentially in spirals may imply that the disk may be in the non-linear phase of an instability.  

However, we take this with the low Toomre~Q (Fig.~\ref{fig:qprof}), high gas velocity dispersion, compact size and large clumps of star formation as holistically fitting a picture that is outlined in theoretical and simulation work describing galaxies in which the Toomre instability drives the internal dynamics of those galaxies. Moreover, the location of the ring is consistent with expectations from Toomre instability theory \citep[e.g.][]{genzel2014}. The ring is located at a radius of $\sim$1~kpc, which is colocated with the rise in Toomre Q at the same radius (Fig.~\ref{fig:qprof}). This is expected in a system in which the galaxy wide instability is driving the flow of gas \citep{genzel2014,dekel2014}.

\vskip 5pt
What Fig.~\ref{fig:tinf} adds is a connection of the gradient in $t_{dep}$ and $\epsilon_{ff}$ directly to the disk instability. Blue-compact disks, that are similar in many properties to IRAS08, are thought to be a critical phase in galaxy evolution \citep{tacchella2016}. Our results suggest that the extreme inflows could sustain extremely high star formation efficiencies, and thus build bulges 20-50$\times$ faster than current prescriptions based on lower assumptions of $\epsilon_{ff}$ than we observe in the center of IRAS08. 

\subsubsection{Outflow as a driver of low depletion times}
Removal of gas via star formation driven winds could lead to an observed decrease in the ratio of SFR to $M_{gas}$. The interpretation of depletion time as the currently observed emission line flux of ionized gas (star formation rate) to the current flux from CO (molecular gas) makes an implicit assumption that the mass of molecular gas is similar to the historic mass, which formed the present population of stars. However, star formation driven winds could reduce the mass of molecular gas in the environment of more extreme star formation. In this case the observed depletion time reflects both the loss of cold gas due to star formation and the loss of cold gas due to outflows, as described in Equation~\ref{eq:tdep}. 

In IRAS08 we observed a lower depletion time in the galaxy center compared to the outer parts. If the mass-loading factor (rate of mass outflow divided by SFR) is higher in the galaxy center than the outskirts, then in principle this could steepen the observed gradient in depletion time and $\epsilon_{ff}$. The mass-loading factor in the galaxy center would need to be of order $\sim20-50\times$ higher in the center for this to completely explain the gradient in $t_{dep}$. \cite{Chisholm2017} observes very fast winds in the  central kiloparsec of IRAS08 ($v_{90}\approx 1000$~km~s$^{-1}$), which would be fast enough to escape the disk easily, but the mass-loading factor is of order $\eta\sim0.05$. This is not sufficient to account for the difference from a disk-value of $t_{dep}\sim1$~Gyr. There is little published work on internal gradients of the mass-loading factor. In simulations, \cite{Kim2020smaug} finds that mass-loading factors are lower for shorter depletion times, which is opposite of the trend needed to explain our results. Our team is using the KCWI data described in this work to measure outflows of ionized gas. Reichardt Chu et al. {\em submitted} finds that there is not strong variation of the mass-outflow rate with $\Sigma_{SFR}$, which would imply that there is likewise not a strong variation with $t_{dep}$. As discussed above, there is a strong correlation of shorter depletion times with higher $\Sigma_{SFR}$. Moreover, they find mass-loading factors of ionised gas that are similar to UV absorption lines, of order unity $\dot{M}_{out}/SFR\sim 1$, which is not sufficient to explain the low $t_{dep}$ in IRAS08. We intend a future paper directly comparing the outflow kinematics to the molecular gas depletion time and gas-mass fraction with the aim of testing models of feedback and star formation regulation. 

\subsubsection{Bars or Mergers as inflow driver}
IRAS08 does not behave similarly in its value nor gradient of $t_{dep}$ to what is observed in either barred disks or merging galaxies (with similar impact parameters and mass-ratios). The comparison of the depletion time of IRAS08 to galaxies with bars and mergers is described in more detail in the Appendix. Here we summarize the results. 

\underline{Merging Galaxies:} Using molecular gas data from the GOALS sample of merging galaxies \citep{Larson2016} we find that similar wide-separation interacting galaxies do not show low global $t_{dep}$. Typically the very low global depletion times are only observed in advanced stages of merging. We also consider the internal gradient in molecular gas depletion times of wide separation mergers. \cite{Espada2018} studies resolved $t_{dep}$ within interacting galaxies with quite similar mass ratios and impact parameters as IRAS08. \cite{Espada2018} finds that  there is much less variation in $t_{dep}$ in the interacting galaxies, than we see in IRAS08, and is shorter at the edge of the galaxy, which is the opposite of IRAS08. Wide-separation,  interactions certainly drive gas inward, but this does not necessarily translate to more efficient star formation in the galaxy center. We also use the data from \cite{Espada2018} to show that in these merging galaxies the depletion times are not consistent with predictions from disk instability theory (Equation 9), and thus satisfying a null hypothesis.  

\underline{Barred Disks:} For bars there is not observational evidence that bars lead to low central $t_{dep}$. Bars are well known to correlate with high central densities of molecular gas \citep{sheth2005,jogee2005,fisher2013}, which are understood theoretically \citep[e.g.][]{athan92,kk04}. However, there is not a well known trend with barred disks have significantly lower $t_{dep}$ in the galaxy center, especially not more than a $\sim$0.2~dex level \citep{Utomo2017}. In IRAS08 there is a central decrease in gas mass surface density (Fig.~\ref{fig:tdep_prof}), which is the opposite of observed gas density profiles in barred disks. Moreover, the bar in IRAS08 is on the small side of typical bars, and the gradient in $t_{dep}$ begins well outside the bar radius. 

It is very important to state the caveat that none of these phenomenon (instabilities, mergers, bars) are mutually exclusive. Simulations now establish that minor-merger style interactions frequently drive the violent disk instabilities \citep{Zolotov2015}. Moreover, rest-frame B-band surveys find that 20-30\% of galaxies at $z\sim1$ are barred \citep{jogee2004}. We note that it is well known that blue-optical surveys significantly underestimate the frequency of bars \citep{eskridge2002}. It would thus imply that significantly more that 30\% of $z\sim1$ galaxies are barred. Indeed, recent studies of observations find bars, spirals and rings are common features in galaxies at $z\sim1.5-3$ \citep{Hodge2019}, and that the historic absence of observations of such features may have been heavily biased by resolution and sensitivity \citep{Yuan2017}. 

\subsection{Comparing IRAS08 to $z\approx1-2$ galaxies}
The observations of IRAS08 we describe in \S3 and \S5 correspond to a star-forming compact rotating galaxy, with a high gas dispersion velocity indicative of a thick disk of very high molecular surface density with low enough Toomre Q to suggest large scale instabilities. This scenario is very similar to the properties of galaxies at $z\sim1-2$ \citep[reviewed in][]{glazebrook2013,Tacconi2020arxiv}. 
Other studies have reached similar conclusions 
\citep{Leitherer2002,Lopez2006,ostlin2009}. The SFR and stellar mass of IRAS08 correspond to those of a main-sequence galaxy at $z\sim1-1.5$. This is also true for the compactness of its 500~nm half-light radius. 

A characteristic feature of galaxies (both main-sequence and bursting) at $z\approx1-2$ is the well known ``clumpy'' star forming regions \citep[e.g.][]{elmegreen2005,genzel2011,guo2015}. \cite{fisher2017mnras} includes IRAS08 in an analysis of local Universe clumpy galaxies from the DYNAMO sample.  The IRAS08 clumps are as bright as 18\% of the total light in H$\alpha$ and several are brighter than 12\%.  This galaxy therefore easily passes quantitative literature definitions of ``clumpy" galaxies \citep[e.g.][]{guo2015,fisher2017mnras}. This similarity to $z\sim1-2$ galaxies, not only in kinematic state (i.e. low-Q and high $\sigma$) but also characteristics of star-forming complexes, suggests that such conditions in a galaxy may facilitate high and/or variable $\epsilon_{ff}$. Moreover, these observations also suggests a connection between this kinematic state and rapid inflows. Given that these same conditions are very common at $z\approx 1-3$, when most star formation in the Universe occurred \citep{madau2014,forster2020} this motivates more study in this area. 

Our results strongly argue for the need for more observations of more turbulent, disk galaxies in which molecular gas and star formation rates can be resolved to scales of $\sim$100~pc. This can only currently be achieved with either local Universe analog samples, like DYNAMO, or lensed galaxies at $z\sim1$ \citep[e.g.][]{Dessauges2019}. Results from such projects would directly inform models of galaxy evolution and possibly make a significant step forward in understanding how bulges form in the early Universe.

\section{Summary}
Our main result is a two order-of-magnitude variation in the molecular gas depletion time and $\epsilon_{ff}$ across a massive blue-compact disk galaxy. We discuss the implications of this for both models of star formation and the evolution of similar blue-compact disks at $z\sim1-3$. 
We find that in the central 50\% of the galaxy typical $\epsilon_{ff}$ values are larger than 10\%, with extreme values as high as 100\%. This variation is much larger than the variation of $t_{ff}$, which accounts for a variable disk thickness. The values and radial variation of $t_{dep}$ and $\epsilon_{ff}$ are very atypical when compared to other disk galaxies in the local Universe \citep{leroy2013,Leroy2017,Utomo2017,Utomo2018,Hirota2018}.

IRAS08 is, however, similar in many properties to the turbulent, compact starbusting disk galaxies of the distant Universe. While highly resolved observations of distant galaxies remain elusive, we can interpret our high spatial resolution observations or IRAS08 as possibly indicating that more efficienct star formation is a common feature of $z\approx 1-2$ galaxy evoution. We show, in Fig.~\ref{fig:sflaw}, that this similarity also translates to a consistency in the resolved Kennicutt-Schmidt relationship between $\Sigma_{SFR}$ and $\Sigma_{mol}$. We find that the relationship between being above the main-sequence and having a steeper $\Sigma_{SFR}-\Sigma_{mol}$ power-law slope is the same at $z\approx0$ as at $z\approx1-2$. This is generically consistent with results showing that galaxies above the main-sequence have short $t_{dep}$ \citep{tacconi2017,Tacconi2020arxiv}, with which our target is also consistent. Our results suggest that the steeper $\Sigma_{SFR}-\Sigma_{mol}$ powerlaw may be driven by a higher $\epsilon_{ff}$ at the cloud scale, as suggested in theory developed in \cite{faucher2013}. Recent observations of molecular clouds in lensed galaxies do suggest higher pressure clouds than what is observed in local spirals \citep[e.g.][]{Dessauges2019}, which may indicate differences in the conversion to stars.

As we have discussed in Section~6, it is hard to reconcile these observations with models assuming constant $\epsilon_{ff}$. Models that do well at describing properties of local spiral galaxies \citep[e.g.][]{ostriker2010,krumholz2012,salim2015} cannot match the observations of IRAS08, or other star-bursting disk galaxies. Theories in which $\epsilon_{ff}$ is variable \citep[e.g.][]{faucher2013} are more consistent. 

We note that for comparing to theory, there is some degeneracy between a truly variable $\epsilon_{ff}$ and a variable feedback-efficiency. If the feedback is more effective at removing molecular gas this could lead to an increase in the observed $\epsilon_{ff}$. \cite{Fisher2019} argues that if the feedback efficiency ($p_*/m_*)$, in Equation~6, were larger in higher $\Sigma_{SFR}$ disk galaxies this could reconcile a number of galaxy properties with locally tested equilibrium star formation theories \citep[e.g.][]{ostriker2010,kim2013}.  
In IRAS08 it is not clear that outflows alone can explain the gradient in $\epsilon_{ff}$. \cite{Chisholm2015} measures very strong winds ($v_{90}\sim1000$~km~s$^{-1}$) in the center of IRAS08. The mass loading factor, however, in the center of the galaxy would need to be of order $\eta\sim 20-50$ to account for the entire decrease $t_{dep}$ below the typical disk value. \cite{Chisholm2017} finds, based on UV-absorption lines, mass-loading factor of 5\% in IRAS08 in photoionized gas. This is for photoionized gas. The relationship between mass-loading factors of different phases is not well-understood, and could be larger in molecular gas, as suggested by \cite{Bolatto2013Natur}. A detailed study of the resolved outflows in IRAS08 is current in progress (Reichardt-Chu {\em in prep}).

We find that the internal properties of IRAS08 are most consistent with a gas inflow being driven by a galaxy wide, violent disk instability \citep[as described in][]{dekel2009,dekel2014}. Not only does IRAS08 exhibit many of the properties similar to those in this theory (e.g. clumpy, high gas velocity dispersion, compact), but we show for the first time, in Fig.~\ref{fig:tinf}, a direct consistency with the prediction from \cite{dekel2014} that in unstable disks the inflow timescale must always be less than $t_{dep}$ in order for an inflow to be maintained. If we interpret IRAS08 as a central burst driven by a violent disk instability, then this galaxy has implications for observations of compaction at high-$z$.  As we show in Fig.~\ref{fig:tdep_prof}, in IRAS08 there is no central pile-up of molecular gas, rather it is exhausted through star-formation on very rapid timescales. Our results imply that high-$z$ blue nugget galaxies could convert their gas very quickly, and make a high-concentration of molecular gas absent. 

We can also consider these extremely rapid growth scenarios in light of red-nuggets at $z>4$ \citep{Glazebrook2017}, and the formation of early-type galaxies. The $\alpha$/Fe abundances of early type galaxies can only be reconciled with their IMFs if they have extremely short formation timescales, in the 10's of Myr \citep{martin2016}. The very short $t_{dep}$ and high $\epsilon_{ff}$ we observed in the galaxy center of IRAS08, is approaching those short times. These $t_{dep}$ are thus not inconsistent with the compact size and short dynamical times of high-z red nuggets, and may provide an avenue to explain the  $\alpha$-enhancement of such galaxies \citep{Kriek2016}.

Ultimately, IRAS08 is only one galaxy. Observations of more galaxies, and observations with alternate methods of measuring star formation efficiency \citep[e.g.][]{Onus2018}, are needed to further confirm this scenario. Moreover,  studies that combine both resolved measurements of $t_{dep}$ with metrics of the feedback, such as outflow kinematics, are needed. Such comparison could determine if the variation in $\epsilon_{ff}$ is due to a true change in star formation efficiency or if feedback is more efficiently removing gas in those regions. We note that to observe sufficiently small spatial scales to measure the $\epsilon_{ff}$ at high $\Sigma_{mol}$ will require either observations of rare local galaxies like ours or lensed galaxies at larger redshift.

\acknowledgments
We are thankful to Cinthya Herrera for help in reducing NOEMA data. This manuscript was significantly improved due to conversations with Adam Leroy, Mark Krumholz and Andreas Burkert, as well as thoughtful comments from the referee. DBF is thankful to Sarah Busch for technical help. DBF acknowledges support from Australian Research Council (ARC)  Future Fellowship FT170100376 and ARC Discovery Program grant DP130101460.  ADB acknowledges partial support form AST1412419. GGK and NMN acknowledges support from ARC DP170103470. 

This work is based on observations carried out under project number W17CB with the IRAM NOEMA Interferometer. IRAM is supported by INSU/CNRS (France), MPG (Germany) and IGN (Spain).

Some of the data presented herein were obtained at the W. M. Keck Observatory, which is operated as a scientific partnership among the California Institute of Technology, the University of California and the National Aeronautics and Space Administration. The Observatory was made possible by the generous financial support of the W. M. Keck Foundation. Observations were supported by Swinburne Keck program 2018A\_W185. The authors wish to recognise and acknowledge the very significant cultural role and reverence that the summit of Maunakea has always had within the indigenous Hawaiian community. We are most fortunate to have the opportunity to conduct observations from this mountain.

\bibliographystyle{yahapj}

\clearpage 
\newpage
\appendix

\section{Wide Separation Interaction \& Possible Galactic Transfer of Gas}

IRAS08 is currently experiencing an interaction with a nearby, lower mass companion galaxy at a separation of $\sim$60~kpc. In interacting systems there is a complex relationship between mass-ratio, interaction distance, gas content and gas depletion timescale \citep[e.g.][]{Combes1994,Renaud2019}. Similar to the expectations from the {\em violent disk instability} we will compare the properties of the interaction in IRAS08 to those observed in other merging galaxies, to determine if IRAS08 exhbits a natural extension of the behavior that is typical of mergers. 

In the GOALS team classification system for mergers \citep{Larson2016} the IRAS08 system is a ``minor merger" due to the mass-ratio. Minor mergers are defined as having a ratio $>$4:1 of the galaxy to the companion that is likely on the early-stage initial approach. IRAS08 has a ratio of at least $\sim$10:1.  Observations indicate that even large mass ratio, distant galaxy interactions can increase the SFR of the larger galaxy \citep{Ellison2008}. Simulations suggest that for those large mass ratios ($\sim$10:1) have only a marginal impact on the structural and kinematic properties of the larger galaxy \citep{cox2008}. 
\begin{figure}
    \centering
    \includegraphics[width=0.5\textwidth]{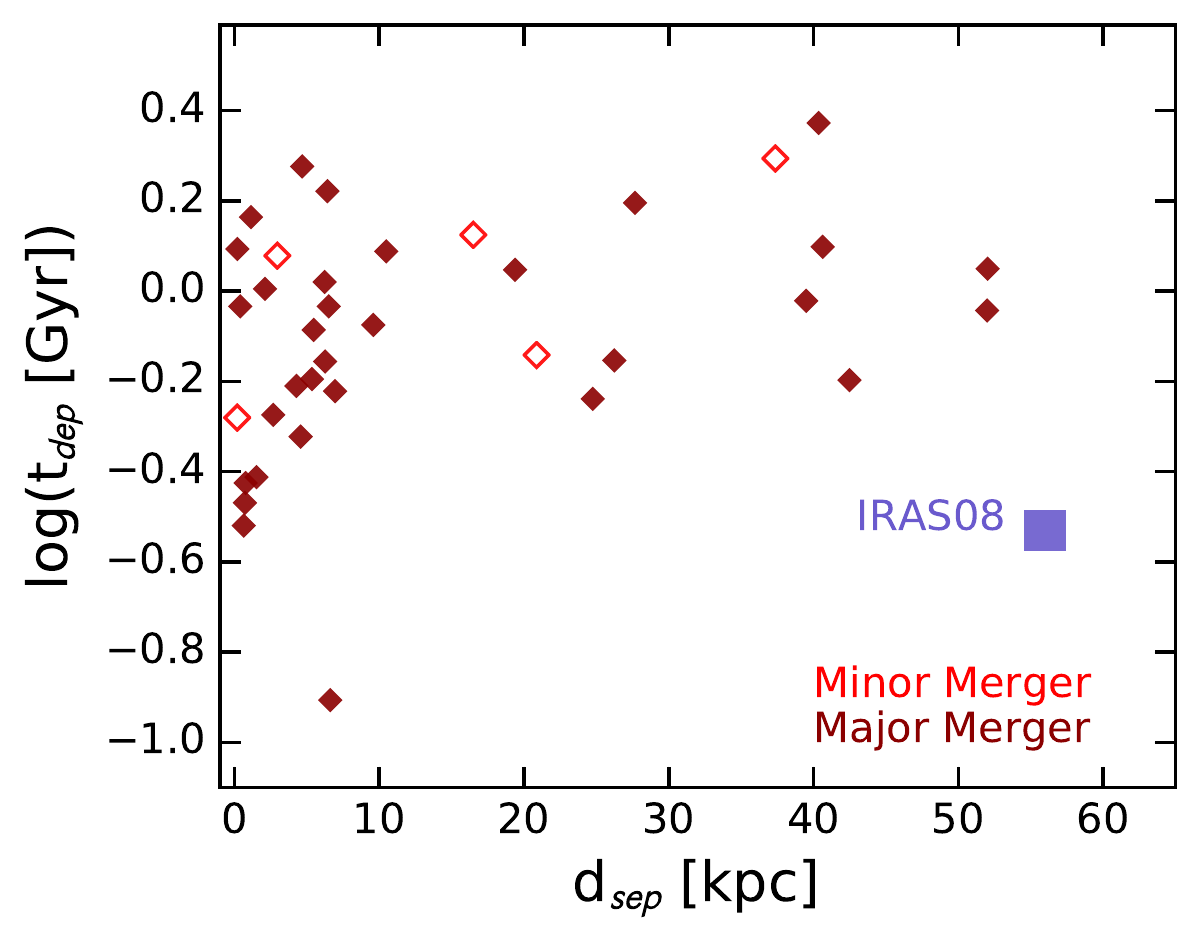}
    \caption{The galaxy averaged molecular gas depletion time of IRAS08 is compared to interacting galaxies from the GOALS sample. Depletion time, $t_{dep}$, is plotted against projected separation of the merging galaxies, $d_{sep}$ for major mergers (mass ratio $<4:1$, filled dark red diamonds), minor mergers (mass ratio $>4:1$, open red diamonds) and IRAS08 (blue square). IRAS08 is a minor merger with a large separation, and a significant outlier from the behavior of merging galaxies in the GOALS sample. }
    \label{fig:goals}
    \centering
    \includegraphics[width=0.75\textwidth]{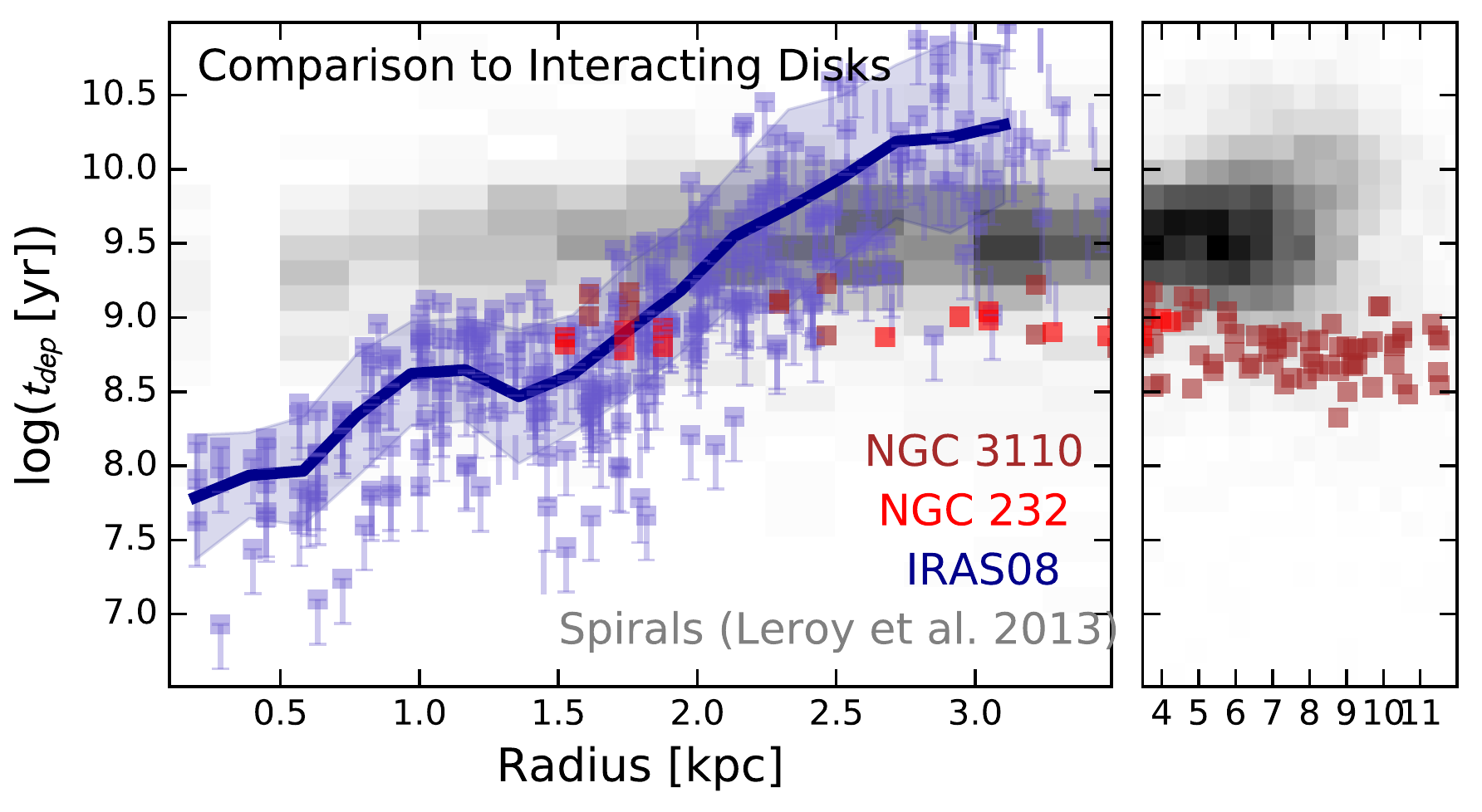}
    \caption{Here we replot Fig.~\ref{fig:tdep_prof} adding wide-seperation early-stage merging galaxies NGC~3110 \& NGC~232 from \cite{Espada2018}. The two interacting systems are shown as brown and red symbols. The separations for these systems are 38~kpc (NGC~3110) and 50~kpc (NGC~232). Neither of these galaxies exhibit similar radial dependence of $t_{dep}$ as IRAS08.
    \label{fig:tdep_prof_mergers}}
\end{figure}

\cite{Cannon2004} show in HI maps that there is a significant reservoir of HI gas extending between the targets. The exact origin of the HI gas is not wholly known. HST/COS observations of IRAS08 show very strong outflows of gas coming out of the center of the galaxy \citep[e.g.][]{Chisholm2015}. Indeed, \cite{Cannon2004} hypothesize that the HI could be related to an outflow. However, based on present observations it is equally likely that the gas has been ejected from the companion, and may represent a transfer of mass from companion to primary galaxy. \cite{Hafen2019arXiv} use simulations to argue that this is one of the most common ways for galaxies to exchange gas. More work is needed to characterise the nature of the large radius HI gas. Independent of its origin the HI plume has a mass of $\sim3\times10^9$~M$_{\odot}$, which is a few percent of the baryonic mass of IRAS08. While by no means a major-merger, this could provide a torquing force to the galaxy. Indeed, in Fig.~\ref{fig:maps} there is a slight asymmetry to the spiral arms that may indicate an asymmetric gravitational potential.  

In Fig.~\ref{fig:goals} we show that the galaxy averaged $t_{dep}$ for IRAS08 is significantly lower than other merging galaxies from the GOALS sample \citep{armus2009}. The GOALS sample is significantly well studied in a large number of publications, with a comprehensive set of observations, and therefore is a useful benchmark for properties of merging galaxies.  \cite{Larson2016} compares morphological merger classification and separation distance to gas content of interacting galaxies. As we show in Fig.~\ref{fig:goals}, significantly low values of $t_{dep}$ is only observed in interacting galaxies with small separations ($d_{sep}<10$~kpc). Even in major-mergers (mass-ratio $<4:1$) with small separation it is not guaranteed that the depletion time is always decreased in merging systems. When viewed as an interacting galaxy IRAS08 is a significant outlier from the typical behavior of interacting galaxies in the GOALS sample, and does not seem to follow the trends of other interacting galaxies. 

Local galaxy M~51 is also experiencing a minor-merger that is far more advanced than IRAS08. The distribution of molecular gas in M~51 is very well studied \cite{leroy2013,Meidt2015,Leroy2017}, and as we show in Fig.~\ref{fig:leffprof} it does not have the same trend of $\epsilon_{ff}$ with radius as in IRAS08.

In Fig.~\ref{fig:tdep_prof_mergers} we show that the radial profile of $t_{dep}$ in two similarly wide-separation interacting galaxies \citep{Espada2018} do not exhibit the same gradient as IRAS08. The two galaxies studied by \cite{Espada2018} are well matched in gas fraction, SFR, total stellar mass, and merger impact parameters to IRAS08, and therefore provide a well controlled comparison. NGC~3110 has a separation of $\sim$40~kpc as mass ratio of 14:1, making it a very early stage minor-merger, like IRAS08. NGC~232 also has a wide-seperation of $\sim$50~kpc, but with a much more significant mass-ratio of 4:5 with its companion. Both galaxies are massive, M$_{star}\approx 6\times10^{10}$~M$_{\odot}$ and star forming $SFR\sim 15-28$~M$_{\odot}$~yr$^{-1}$. Neither galaxy shows the same strong decline in $t_{dep}$ toward the galaxy center. Indeed, both NGC~3110 and NGC~232 show a mild increase in $t_{dep}$ in the galaxy center. Other significant differences exists between these two interacting systems and IRAS08. Both NGC~3110 and NGC~232 have very strong gradients in molecular gas velocity dispersion, and both have a disk averaged surface density that is much lower, $\Sigma_{mol}^{disk}\approx 8-21$~M$_{\odot}$~pc$^{-2}$. Whereas IRAS08 has a nearly constant $\sigma(R_{gal})$, Fig.~\ref{fig:sig_prof}, and the surface density at the edge of the disk of order $\sim$100~M$_{\odot}$~pc$^{-2}$.  \cite{Espada2018} shows that the molecular gas surface density in both NGC~232 \& NGC~3110 is high in the center, and provide arguments that this is driven by the interaction. However, in NGC~232 and NGC~3110 this does not translate to a lower $t_{dep}$ in the galaxy center. 

We can also find that NGC~232 and NGC~3110 do not satisfy predictions of the violent disk instability model. \cite{Espada2018} used numerical simulations to established that the gas flows in NGC~232 and NGC~3110 are most likely due to the interaction from the host. These systems therefore offer a good test to determine of the violent disk instability model. If these galaxies have low $t_{inflow}$ and low $Q$ then this would weaken the case that these metrics are meaningful for IRAS08. We consider the disk and central values for these galaxies using data from \cite{Espada2018} as inputs into Equations~3 \& 4 of this paper. At large radius, in the disk, we find $Q\approx 2.9$~\&~1.75 for NGC~232 and NGC~3110, respectively. We also find inflow timescales of 7~Gyr and 12~Gyr. In both cases the Toomre parameter suggests the disk is stable and $t_{inflow}>t_{dep}$. This is not consistent with violent disk instabilities as drivers of the gas inflow in NGC~232 and NGC~3110. Even in the galaxy center the value for Toomre Q remains high; for both targets $Q(R=1~kpc)\approx 2$. To be clear, this does not mean that IRAS08 is necessarily driven by the instability. It is however a useful to see that $t_{inflow}$ and $Q$ do not result in ``false positive" results when we have independent evidence that a gas flow is not driven by an instability.  

If the gas flow in IRAS08 is driven by its interaction, then this interaction would be different from other interactions observed in the local Universe. IRAS08 has a molecular gas depletion time that is as short as what is observed in advanced stage mergers \citep{Wilson2019,Bemis2019}, yet has completely different morphology and kinematics from those systems. Moreover, the strong gradient in $t_{dep}$ is not observed in other wide-separation interacting systems \citep{Espada2018}.  

\section{Stellar Bar}
The F550M image of IRAS08 shows a stellar bar (Fig.~\ref{fig:maps}) in the center of this galaxy. Simulations clearly establish that bars can impact the distribution of gas in galaxies \citep[e.g.][]{athan92,simkin1980}. Here we consider the possibility that this bar may contribute to the radial change in $t_{dep}$ and $\epsilon_{ff}$.

The exact impact of bars on gaseous disks is somewhat complex. Along bars star formation tends to be suppressed. This is thought to be due to strong shocks \citep{athan92}, which increase the velocity dispersion of the molecular gas \citep{Maeda2018}. The gas then concentrates in the galaxy centers. Indeed, observations of barred galaxies show preferentially higher molecular gas mass surface densities than non-barred galaxies \citep{sheth2005,jogee2005,fisher2013}. However, observations do not show strong evidence for a significant change in the $t_{dep}$ inside of bars \citep{fisher2013}. Indeed, the majority of disk galaxies in the HERACLES survey \citep{leroy2013} are barred systems, as it draws from the general population of star forming disk galaxies. It is this sample that we use for comparison of the radial gradient in $t_{dep}$ (Fig.~\ref{fig:tdep_prof}). If barred galaxies showed a significant decrease in $t_{dep}$ in the central kiloparsec this would be dected in Fig.~\ref{fig:tdep_prof}, but we see only a slight change toward the center. 

It is important to point out that there are substantial differences between the properties of galaxies in most simulations of barred galaxies and in IRAS08 \citep[e.g.][]{athan92,regan1997,Maciejewski2002}. The overall gas velocity dispersion and total gas fraction in IRAS08 are significantly larger than in simulations set to match the Milky Way. Moreover, the bar in IRAS08 is only $\sim$2$\times$ larger that than the Toomre length derived for this galaxy, which gives an expected size of molecular clouds. Whereas in more typical local barred galaxies the characteristic giant molecular cloud size is 10-100$\times$ smaller than the bar. It is not clear how this might affect the interaction between bars and gas. For example, \cite{Maciejewski2002} find that small-scale bars do not produce shocks in galaxy centers, a similar phenomenon could occur in IRAS08. To our knowledge there are no simulations of the impact of bars in a gas medium that has a high velocity dispersion.
\begin{figure}
    \centering
    \includegraphics[width=0.48\textwidth]{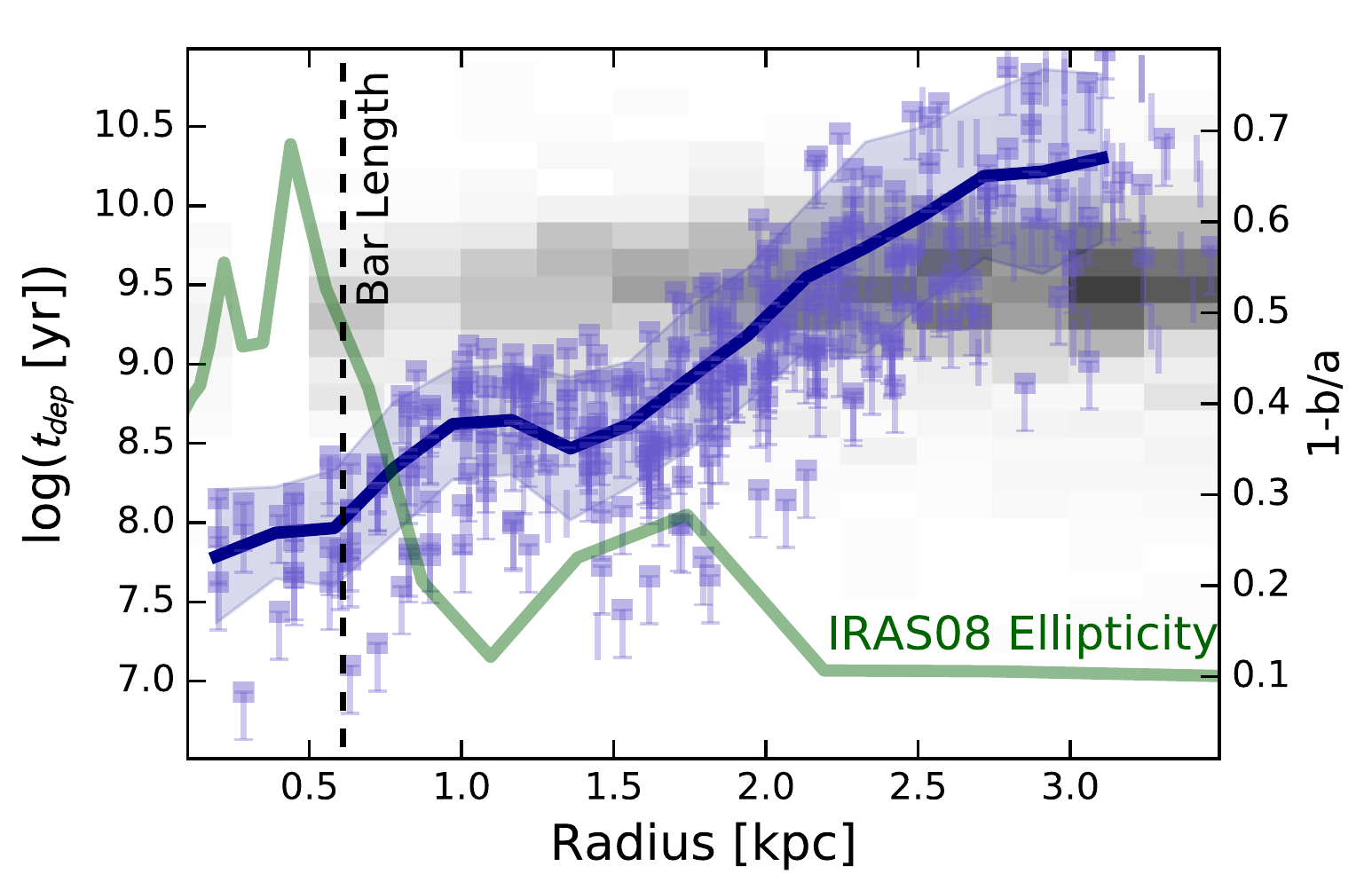}
    \caption{The above figure compares the radial profile of $t_{dep}$ to the ellipticity of the 500~nm flux. The rise in ellipticity of starlight indicates the location of the bar. The bar in IRAS08 likely only impacts the gas at radii $R_{gas}<1$~kpc, which is much smaller than the bulk trend in $t_{dep}$. }
    \label{fig:bar}
\end{figure}

In Fig.~\ref{fig:bar} we compare the ellipticity of the star-light to the molecular gas depletion time, both as function of radius within the galaxy. There is debate in the literature about exactly where to place the bar length \citep[see discussion in][]{erwin2005,marinova2007}. We use the ellipticity profile of IRAS08 to identify the bar, and choose the bar length as the radius beyond which the ellipticity decrease by 15\% from the peak value. Numerical simulations find that this radius is in good agreement with bar lengths as defined by orbital analysis \citep{Martinez2006}. Moreover, when overlaying this ellipse on the F550M image we find that this radius corresponds to the point at which the bar meets the ring of gas. Fig.~\ref{fig:bar} shows that the decrease in $t_{dep}$ begins far beyond the radius of the bar. 

We note that the ring just beyond the bar, identified as a minimum in the ellipticity profile, is colocated with a relative increase in $t_{dep}$ and a decrease in $\epsilon_{ff}$. Typically, in nearby spiral galaxies rings are associated with more efficient star formation than in the surrounding disk \citep[e.g.][]{kk04}. Rings are found to be very common in disk galaxies at $z>1$ \citep{genzel2014}, and 4 of 10 galaxies in the DYNAMO sample of gas-rich, clumpy disks show evidence of rings \citep{fisher2017mnras}. If rings behave differently in gas-rich galaxies than in local spirals this may be an interesting avenue for further research. 

In general, there is not strong evidence that bars lead to enhanced star formation efficiencies in their centers. In IRAS08 the general trend of decreasing $t_{dep}$ begins at radii 3-4$\times$ the bar radius, also suggesting this may be a galaxy wide phenomena rather than the bar.  

Empirically speaking it is very difficult to determine if the presence of the bar should impact our treatment of IRAS08 as a similar phenomenon as $z>1$ unstable disk galaxies. First, the bar length in IRAS08 is $\lesssim1$~kpc. If IRAS08 were observed at the same resolution as a $z\sim2$ galaxy with HST this bar would be covered by only 1-2 resolution elements. It would thus not be so straightforward to identify the bar. Moreover, bars are far more easy to identify in redder wavelengths \citep{eskridge2002}, and observations of restframe V-band light have significantly lower signal-to-noise at $z=2$, it is conceivable that systematic uncertainties in observations lead to a lower frequency of observed bars at $z>1$. Finally, bars, rings and spiral structure in local Universe disks are known to be related phenomena \citep[for review][]{kk04}. Sprials and rings are by no means absent from the high-$z$ universe. Spiral galaxies have been observed at $z>2$ \citep{Yuan2017}, and recent work with ALMA identifies central concentrations of elongated structures in galaxies at $z>2$ \citep{Hodge2019}. \cite{inoue2016} argues that massive star forming clumps in many galaxies at $z>1$ may be intrinsically linked to spirals. As noted above, rings are likewise very common at $z>1$. In short, we will not know if small scale bars (R$\lesssim$1~kpc) are common at $z>1$ until the advent of next-generation adaptive optics instruments such as VLT-MAVIS, ELT-MICADO, or TMT-NFIRAOS come on-line in the later part of the next decade. 

In summary though bars are well known to drive high molecular gas mass surface density in galaxy centers, the evidence from samples of barred galaxies is that they do not lead to significant changes to the gradient of $t_{dep}$. Moreover, in IRAS08 the bar is quite small, whereas the gradient in $t_{dep}$ and $\epsilon_{ff}$ is a phenomenon that covers the entire disk. At a finer detail, the bar in IRAS08 is probably playing some role in gas redistribution in the central $R_{gas}<0.5$~kpc, however, it does not seem to be the main-driver of the full gradient in $t_{dep}$.

\end{document}